\documentclass{aastex63}

\usepackage{CJK}
\received{X, 2022}
\revised{X, 2022}
\accepted{}%\today

\submitjournal{ApJ}

\shorttitle{Mass-Ratio Distribution of Binaries From the LAMOST-MRS Survey}
\shortauthors{Li et al.}

\begin{document}
   \begin{CJK}{UTF8}{gbsn}

   \title{Mass-Ratio Distribution of Binaries From the LAMOST-MRS Survey\footnote{Released on , , 2022}}

\correspondingauthor{Zhanwen Han, Chao Liu}
\email{zhanwenhan@ynao.ac.cn, liuchao@bao.ac.cn}

\author[0000-0003-3832-8864]{Jiangdan Li}
\affiliation{Yunnan Observatories, Chinese Academy of Sciences (CAS), 396 Yangfangwang, Guandu District, Kunming 650216, P.R. China}
\affiliation{Key Laboratory for the Structure and Evolution of Celestial Objects, CAS, Kunming 650216, P.R. China}
\affiliation{University of Chinese Academy of Sciences, Beijing 100049, China}

\author[0000-0002-2577-1990]{Jiao Li}
\affiliation{Yunnan Observatories, Chinese Academy of Sciences (CAS), 396 Yangfangwang, Guandu District, Kunming 650216, P.R. China}
\affiliation{Key Laboratory for the Structure and Evolution of Celestial Objects, CAS, Kunming 650216, P.R. China}
\affiliation{Key Lab of Space Astronomy and Technology, National Astronomical Observatories, Chinese Academy of Sciences, Beijing 100101, China}

\author[0000-0002-1802-6917]{Chao Liu}
\affiliation{University of Chinese Academy of Sciences, Beijing 100049, China}
\affiliation{Key Lab of Space Astronomy and Technology, National Astronomical Observatories, Chinese Academy of Sciences, Beijing 100101, China}

\author[0000-0002-6647-3957]{Chunqian Li}
\affiliation{University of Chinese Academy of Sciences, Beijing 100049, China}
\affiliation{CAS Key Laboratory of Optical Astronomy, National Astronomical Observatories, Chinese Academy of Sciences, Beijing 100101, China}

\author{Yanjun Guo}
\affiliation{Yunnan Observatories, Chinese Academy of Sciences (CAS), 396 Yangfangwang, Guandu District, Kunming 650216, P.R. China}
\affiliation{Key Laboratory for the Structure and Evolution of Celestial Objects, CAS, Kunming 650216, P.R. China}
\affiliation{University of Chinese Academy of Sciences, Beijing 100049, China}

\author{Luqian Wang}
\affiliation{Yunnan Observatories, Chinese Academy of Sciences (CAS), 396 Yangfangwang, Guandu District, Kunming 650216, P.R. China}

\author{Xuefei Chen}
\affiliation{Yunnan Observatories, Chinese Academy of Sciences (CAS), 396 Yangfangwang, Guandu District, Kunming 650216, P.R. China}
\affiliation{Key Laboratory for the Structure and Evolution of Celestial Objects, CAS, Kunming 650216, P.R. China}

\author{Lifeng Xing}
\affiliation{Yunnan Observatories, Chinese Academy of Sciences (CAS), 396 Yangfangwang, Guandu District, Kunming 650216, P.R. China}

\author{Yonghui Hou}
\affiliation{Nanjing Institute of Astronomical Optics \& Technology, National Astronomical Observatories, Chinese Academy of Sciences, Nanjing 210042, China}
\affiliation{School of Astronomy and Space Science，University of Chinese Academy of Sciences}

\author{Zhanwen Han}
\affiliation{Yunnan Observatories, Chinese Academy of Sciences (CAS), 396 Yangfangwang, Guandu District, Kunming 650216, P.R. China}
\affiliation{Key Laboratory for the Structure and Evolution of Celestial Objects, CAS, Kunming 650216, P.R. China}
\affiliation{University of Chinese Academy of Sciences, Beijing 100049, China}

\begin{abstract}

Binary evolution leads to the formation of important objects crucial to the development of astrophysics, but the statistical properties of binary populations are still poorly understood. The LAMOST-MRS has provided a large sample of stars to study the properties of binary populations, especially for the mass ratio distributions and the binary fractions. We have devised a Peak Amplitude Ratio (PAR) approach to derive the mass ratio of a binary system based on results obtained from its spectrum. By computing a cross-correlation function (CCF), we established a relationship between the derived mass ratio and the PARs of the binary systems. By utilizing spectral observations obtained from LAMSOT DR6 \& DR7, we applied the PAR approach to form distributions of the derived mass ratio of the binary systems to the spectral types. We selected the mass ratio within the range of $0.6-1.0$ for investigating the mass-ratio distribution. Through a power-law fitting, we obtained the power index $\gamma$ values of $-0.42\pm0.27$, $0.03\pm0.12$, and $2.12\pm0.19$ for A-, F-, and G-type stars identified in the sample, respectively. The derived $\gamma$-values display an increasing trend toward lower primary star masses, and G-type binaries tend to be more in twins. The close binary fractions (for $P\lesssim 150\,{\rm d}$ and $q\gtrsim 0.6$) in our sample for A, F and G binaries are $7.6\pm 0.5 \%$, $4.9\pm 0.2 \%$ and $3.7 \pm 0.1 \%$, respectively. Note that the PAR approach 
%works well for main sequence binaries with orbital periods less than $\sim 150$ days and mass ratios between $0.6$ and $1.0$, and this approach 
can be applied to large spectroscopic surveys of stars.
%The mass-ratio distribution in our study can be explained by the current binary formation mechanisms.

\end{abstract}


\section{INTRODUCTION}
\label{sect:intro}

Most of what we know about the Universe comes from stars, and about half of the stars are in binaries. Early-type stars tend to be more likely found in pairs, and a binary fraction of 70\% is reported by \citet{Sana2012} for massive stars. The binary fraction drops down to $\sim44\%$ for Solar-type stars but still suggests a non-negligible effect in the binary population \citep{Duchene2013}.
The interactions between the component stars of binaries change their destiny \citep{Jones2017}. Binary evolution is much more complicated, and results in the formation of stellar objects with exotic observational phenomena. Binary evolution produces type Ia supernovae, double black holes, double neutron stars, millisecond pulsars, X-ray binaries, etc.\ and these compact systems contribute to the chemical evolution of galaxies and the spectral energy distribution at short wavelength of early type galaxies, and provides re-ionizing photons of the early Universe (for a review, see \citealt{Han2020}.) 

Observational properties of binary populations are basic inputs for many studies, e.g. gravitational wave sources and progenitors of type-Ia supernovae. Observational properties of binary populations, including binary fraction, binary orbital period distribution, mass-ratio distribution, orbital eccentricity distribution, and the dependence of the distribution on stellar type and metallicity, are essential inputs for stellar evolution. These properties enable us to understand better the binary evolution, the formation of binary-related objects, galactic evolution, and star formation.
For example, if an observed mass-ratio distribution is shown to be consistent with the one in which the two components follow the same initial mass function (IMF), the component stars may form independently with random pairings \citep{Abt1990, Tout1991, McDonald1995}. Alternatively, co-related component masses of close binaries indicate coevolution of the two components during their pre-MS phase via physical processes such as fragmentation, fission, competitive accretion, and/ or mass transfer (MT) through Roche-lobe overflow (RLOF) \citep{Kroupa1995b, Kroupa1995a, Bate1997, Kouwenhoven2009}.  

Much observational effort has been made, but the statistical properties of binaries are still unclear. Mass-ratio distributions and their dependence on stellar parameters or environments (different metallicities or galactic disk/halo) have not been understood yet. Even for the most observable solar-type binaries, the mass-ratio distribution is controversial. In earlier studies, the two components in a binary system are found to have very different masses \citep{Duquennoy1991} and the mass-ratio distribution has an increasing trend toward small secondary masses, while \citet{Raghavan2010} claimed that the distribution is more or less uniform when the mass ratio is less than 0.9. Recent studies take that the mass-ratio distribution is strongly correlated with the primary mass and can be described as a power law \citep{Moe2017}. On the other hand, mass ratio determines whether mass-transfer is stable or not in a binary system, and consequently the binary system will undergo a stable mass transfer or experience a common envelope evolution, leading to the formation of different types of objects. Such a process is difficult to tackle with a theoretical approach due to its complexity and observational constraints are needed (see, however, \citealt{Ge2010, Ge2015, Ge2020b,Ge2020a} for recent progress).  Mass-ratio distribution needs to be investigated further. 

In recent years, large photometric surveys and spectroscopic surveys have been carried out, and a huge amount of data are released, leading to an increasing number of studies on binary stars. Eclipsing binaries can be analyzed from light curves from photometric surveys, e.g.\ the Panoramic Survey Telescope and Rapid Response System \citep{Flewelling2016}, and the Zwicky Transient Facility \citep{Bellm2019a, Graham2019} etc. Properties of the binaries, including mass ratios, can be obtained from the analysis, but most of the binaries analyzed are in short orbital periods ($\lesssim10 {\rm d}$), since binaries with longer orbital periods are less likely to have eclipses. Spectroscopic data are much less affected by the selection effect of orbital inclinations and therefore can provide information on binaries with longer orbital periods.  Examples of investigation on binaries from large spectroscopic surveys are \citet{Liu2020} from the Large Sky Area Multi-Object Fiber Spectroscopic Telescope (LAMOST) survey and \citet{Majewski2017} from the Apache Point Observatory Galactic Evolution Experiment. However, the spectroscopic binary sample with known properties is still much smaller than that of photometric one.  The number of the photometric binaries with properties derived is about 100\,000, while the number of spectroscopic binaries is around 10,000 in the Ninth Catalogue of Spectroscopic Binary Orbits \citep{Pourbaix2004} and the Geneva-Copenhagen Survey Catalogue \citep{Nordstrom2004, Holmberg2009}.

Recent spectroscopic surveys provided a large amount of double line spectroscopic binaries (SB2). For example, \citet{Matijevic2010} found 123 SB2s in the Radial Velocity Experiment Wide-field spectroscopic surveys of the stellar content of the Galaxy (RAVE), \citet{Merle2017} detected 342 SB2s from the Gaia-ESO survey, \citet{El-Badry2018} found around 2\,500 SB2 candidates in Apache Point Observatory Galactic Evolution Experiment (APOGEE) DR13, \citet{Traven2020} obtained 12\,760 FGK SB2s with stellar properties from the GALactic Archaeology with HERMES (GALAH) survey, and \citet{Kounkel2021} showed 7\,273 candidate SB2s in APOGEE DR16 and DR17. A five-year LAMOST medium resolution survey (LAMOST-MRS) has been carried out since 2018. The survey is designed to observe 2 million stellar spectra with a resolution of $R\sim 7\,500$ and a limiting magnitude of $G \sim 15$ mag during bright/gray nights \citep{Liu2020}. With the medium resolution spectra from LAMOST Data Release (DR) 6 and DR 7, \cite{Lichunqian2021} obtained 3\,133 SB2 candidates, 95\% of which are newly discovered. The method that \cite{Lichunqian2021} developed to search for SB2s is based on using a Cross-Correlation Function (CCF) and applying a Gaussian smooth to measure the radial velocity (RV) of the sample stars. \cite{Lichunqian2021} have adopted a Cross-Correlation Function (CCF) approach and applied a Gaussian smooth technique to the observed spectra to measure their radial velocity (RV) to search for SB2s from the LAMOST database. To find an SB2, we need to see the spectra of the two components simultaneously. The luminosities of the two components are required to be comparable and the radial velocity difference ($\Delta RV$) is shown as double peak is large enough. The luminosity difference and the RV difference of a binary depend on its mass ratio, and consequently, we can derive mass ratio information for the detected SB2 candidates using the LAMOST data set, which is the purpose of this paper. It is ideal for identifying SB2 when the individual spectral components appear in the observations simultaneously and the component stars display comparable luminosity and significant RV variations. Motivated by the large sample of SB2 recently identified from \citet{Lichunqian2021}, in this work, we aim to investigate the dependence of the luminosity difference and RV variations on the mass ratio of the SB2 systems.

In this paper, we study the mass-ratio distribution of binaries from the SB2 candidates \citep{Lichunqian2021}. The paper is organized as follows. We briefly describe the LAMOST data in Section~\ref{sec:data},  and the methods and models used in Section~\ref{sec:method}. Results are shown in Section~\ref{sec:results}. Discussions and conclusions are given in Section~\ref{sec:discussion} and~\ref{sec:conclusion}, respectively.

\section{The data}
\label{sec:data}

We adopt the observational results using spectra of LAMOST-MRS from \cite{Lichunqian2021} to perform this mass ratio distribution. LAMOST is a special quasi-meridian reflecting Schmidt telescope located in Xinglong Station of National Astronomical Observatory, China.
The telescope has 4\,000 fibers installed on a 5-degree-FoV focal plane and can obtain spectra of 4\,000 celestial objects simultaneously \citep{Cui2012, Zhao2012}. The pilot and the 1st stage LAMOST survey were made using spectrographs with a $R\sim 1\,700$ from 2011 to 2017. 
The 2nd stage survey, i.e. LAMOST-II, contains both low resolution ($R\sim 1\,700$) survey and medium resolution ($R\sim 7\,500$) survey (LAMOST-MRS), started in 2017. See \citet{Liu2020} for a detailed review of the LAMOST-MRS survey. 

\begin{figure}
\plotone{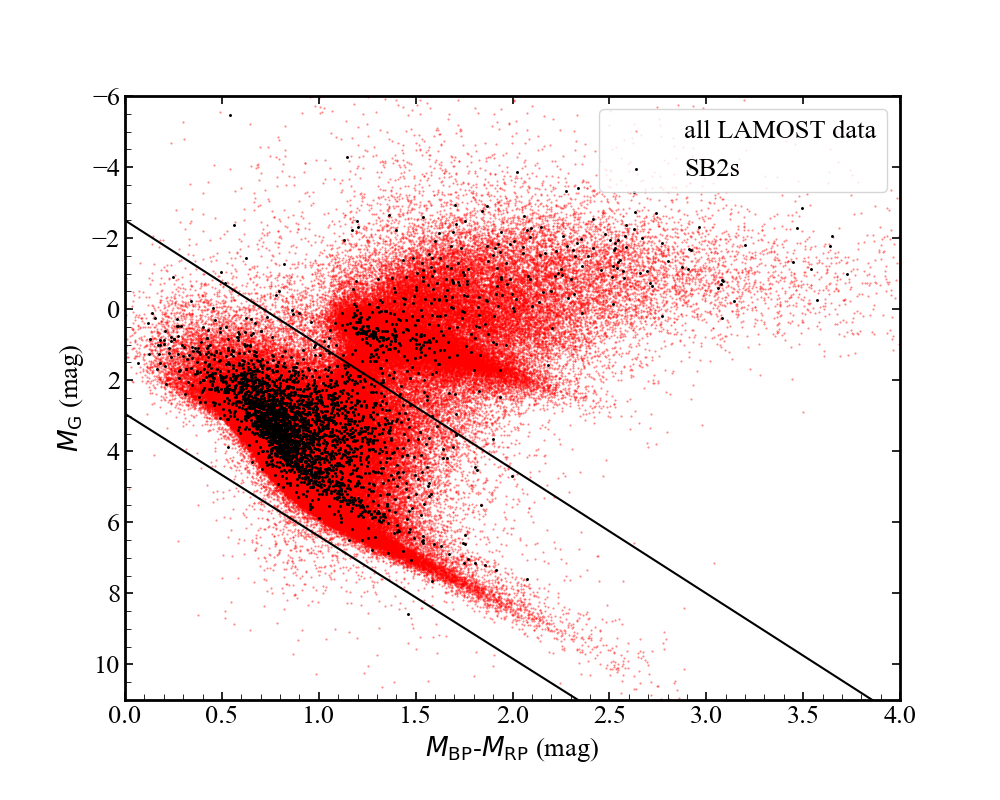}
\caption{The color-magnitude diagram of 3\,133 SB2 candidate stars (black dots) identified from \cite{Lichunqian2021} and LAMOST-MRS DR6 \& DR7 data (red dots). The photometric measurements were collected from {\it Gaia} DR2, the color difference between the blue band ($M_{\mathrm BP}$) and the red band ($M_{\mathrm RP}$) are shown on the x-axis and the absolute magnitudes in G-band ($M_{\mathrm{G}}$) are plotted in the y-axis. Our final sample of MS stars was selected based on their locations on the diagram between the two solid lines.
}
\label{fig:HRdiagram}

\end{figure}

Adopting a cross-correlation technique, \cite{Lichunqian2021} reported detection of 3\,133 SB2 candidates from a total number of 952\,747 stars in LAMOST-MRS DR6 and DR7. 
We selected the candidates with an error less than 10$\%$ for their peak magnitudes from the SB2 sample. We then cross-matched the selected SB2 candidates with {\it Gaia} DR2 to collect their G-band, Bp-band, and Rp-band magnitudes \citep{GaiaCollaboration2016, GaiaCollaboration2018, Arenou2018}. We cross-matched our spectra with {\it Gaia} using a 3 arcsec searching cone because the LAMOST fibers have a diameter of 3.3 arcsec \citep{Cui2012}. We show the color-magnitude diagram of SB2 candidates (black dots) in Figure~\ref{fig:HRdiagram}. No interstellar reddening corrections were applied to the photometric measurement because Main Sequence (MS) stars have negligible reddening. We visually inspected the photometric measurements of stars and placed two lines on the diagram that are symbolic of the position of the zero-age MS (lower line) and terminal-age MS (upper line). We then selected MS binaries with locations between these two lines shown on the color-magnitude diagram in Figure~\ref{fig:HRdiagram}. 

We then classified the MS sample binaries into different spectral types following the Morgan-Keenan (MK) classification by using the effective temperatures given by LAMOST Stellar Parameter (LASP) pipeline \citep{Wu2011, Wu2014, Luo2015}. LASP was developed to extract stellar parameters for LAMOST spectra observed in both Low-Resolution Survey (LRS) and MRS. It employs the ULySS software \citep{Koleva2009} to analyze the spectra and obtains the effective temperature by minimizing the $\chi^{2}$ values between the observed spectrum and a model spectrum generated by an interpolator built based on the ELODIE library \citep{Prugniel2001, Prugniel2007}. Only temperatures below 8\,500\,K are given because the spectra for A-type stars are insensible to their temperatures. The errors of estimated effective temperature have values typically less than 100\,K \citep{Wang2020}.

In Table~\ref{tab:1}, we list the number of stars in LAMOST and the number of SB2 candidates for each spectral type. Considering the fact that the effective temperature of late-type stars assigned by LASP is unreliable, we discarded the stars with temperatures below 5\,150\,K. We also exclude the stars with temperatures above 8\,000\,K to decrease the possibility of early-type stars mixing in the sample. As a result, we have SB2 candidates of spectral type A, F, and G. We also show the SB2 numbers within 500, 1\,000, and 2\,000\,${\rm pc}$ of the Sun in the Table~\ref{tab:1}.
Our final sample includes a collection of $2\,003$ SB2 candidate stars from 149\,456 MS stars.

\begin{deluxetable*}{ccccccc}
\tablenum{1}
\tablecaption{The number of SB2 candidates identified for each spectral types (SPEC)\label{tab:1}}
\tablewidth{0pt}
\tablehead{
\colhead{$T_{\mathrm{eff}}$(K)} & \colhead{SPEC}  & \colhead{$N_{\mathrm{ALL}}$}& \colhead{$N_{\mathrm{SB2}}$} & \colhead{$N_{\mathrm{SB2}}(500\,\mathrm{pc})$} & \colhead{$N_{\mathrm{SB2}}(1\,000\,\mathrm{pc})$} & \colhead{$N_{\mathrm{SB2}}(2\,000\,\mathrm{pc})$} 
}
\startdata
7\,200-8\,000 & A & 10\,228  & $226$    & $29 $ & $133 $  & $211 $\\
6\,000-7\,200 & F & 60\,231  & $882$    & $183$ & $639 $   & $860 $\\
5\,150-6\,000 & G & 78\,997  & $895$    & $311$ & $738 $   & $870 $\\
Total       & All & 149\,456 & $2\,003$ & $523$ & $1\,510$ & $1\,941$ \\
\enddata
\tablecomments{$N_{\mathrm{SB2}}$ is the number of SB2 candidates from the LAMOST-MRS DR6 \& DR7. The candidates are divided into 3 parts, e.g. spectral types, based upon their $T_{\mathrm{eff}}$ ranges. $N_{\mathrm{SB2}}(500\,\mathrm{pc})$, $N_{\mathrm{SB2}}(1\,000\,\mathrm{pc})$ and $N_{\mathrm{SB2}}\,(2\,000\,\mathrm{pc})$ show the number of SB2 candidates within 500, 1\,000 and 2\,000\,{\rm pc} from the Sun, respectively. }
\end{deluxetable*}

\section{Method}
\label{sec:method}

In this section, we devise an approach to derive the mass ratios of SB2 binaries from LAMOST-MRS data. We construct a mock sample of binaries, obtain their synthetic spectra, and apply the CCF method of \cite{Lichunqian2021} to detect SB2s from the synthetic spectra. SB2s show two peaks in their CCF. We found that there exists a relation between the PARs and the mass ratios for SB2s. Applying the relation, we can derive the mass ratios and consequently their distribution from the PARs of the observed SB2s of LAMOST-MRS.

\subsection{The CCF method of Li et al. (2021)}
\label{sec:examination}

\cite{Lichunqian2021} developed an automatic method to search for binaries using observations from LAMOST-MRS data. The method is based on calculating the Cross-Correlation Function (CCF) between a star's spectrum and a template spectrum. They applied a template matching and a Gaussian smoothing to find the number, the location, and the amplitudes of peaks of the CCF.
The template \cite{Lichunqian2021} adopted is the observed solar spectrum with a resolution of 100\,000  \citep{Kurucz1984}. They have altogether found 3\,133 SB2  candidates from 952\,747 blue arm spectra with SNR$\geq10$ from LAMOST-MRS. Above 95\% of the candidates are new detections. 
The authors showed that radial velocity difference between the two components of a binary system ($\Delta RV=|RV_1-RV_2|$) is the key factor for SB2 candidate identification. The lower limit of $\Delta RV$ is found to be $\sim 50$\,km/s, which mainly depends on the resolution of the observed spectra.

\subsection{A PAR approach for the derivation of mass ratios}
\label{sec:PAR}
The CCF approach of \cite{Lichunqian2021} identifies a system to be a binary if there are two peaks in its CCF, and detection mainly depends on the radial velocity differences of the SB2s. However, as can be imagined, the amplitudes of the peaks should depend on the mass ratio of the secondary to the primary ($q={m_2 / m_1}$). 
To test the hypothesis, we constructed synthetic spectra for a binary grid with different temperatures, mass ratios, RV differences, and signal-to-noise ratio (SNR) levels. We described the procedures as follows.

\begin{figure}
\plotone{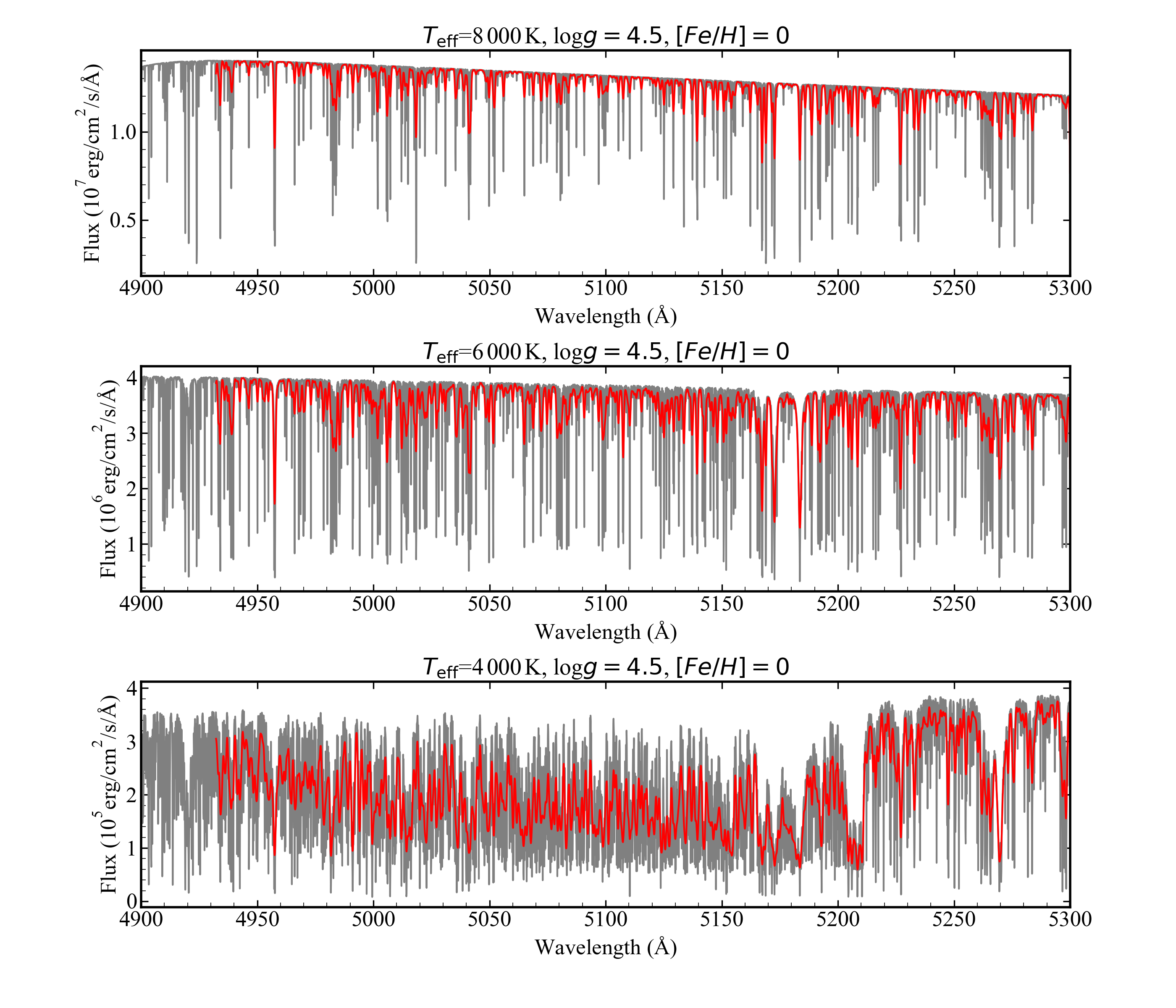}
\caption{The model spectra with $T_{\mathrm{eff}}$ values of 8\,000\,K ( top panel), 6\,000\,K (middle panel) and 4\,000\,K (bottom panel), the ${\mathrm{log}}g$ is set to be 4.5 and metallicity is set to be solar.  The grey lines are the original spectra with a resolution of 51\,000, the red lines are the degraded spectra with a resolution of 7\,500.
\label{fig:spectra}
}
\end{figure}

We first employed the radiative transfer code {\it SPECTRUM} \citep{Gray1994} to generate a grid of synthetic stellar spectra from the $ATLAS$9 stellar atmospheric model \citep{Kurucz1993}. The model is for the plane-parallel atmosphere with the assumptions of local thermodynamic equilibrium (LTE). The model spectra have a range of effective temperatures between 4\,000\,K and 8\,000\,K, corresponding to K-, G-, F-, and A-type stars, with a step size of 100\,K. We adopted a standard value of surface gravity for MS star with ${\mathrm{log}}g$=4.5 ($g$ is in ${\rm cm\,s^{-2}}$), 
and the metallicities [Fe/H] are set to be 0, i.e. a solar metallicity. We constructed the model spectra covering the wavelength regime from 4\,900 {\AA} to 6\,900 {\AA}, and downgraded the resolution from 51\,000 to 7\,500 to match that of the LAMOST-MRS spectra via convolution with the $laspec$ code \citep{Zhang2020, Zhang2021}. 
In Figure~\ref{fig:spectra}, we show three examples of model spectra constructed from the ATLAS9 grids. In the top panel, we plotted the original high resolution model spectrum (R $\sim$ 51\,000) with $T_{\mathrm{eff}}$= 8\,000\,K, ${\mathrm{log}}g$ = 4.5, and [M/H] = 0 in gray, and overplotted the degraded spectrum with R $\sim$ 7\,500 in red. The model spectra constructed with $T_{\mathrm{eff}}$= 6\,000\,K and $T_{\mathrm{eff}}$= 4\,000\,K with the same ${\mathrm{log}}g$ and [M/H] are shown in the middle and the bottom panels, respectively. 

We then constructed a grid of synthetic spectra for MS binaries of solar metallicities from the $ALTAS$9 stellar spectra above. 
For an MS binary system, both components are MS stars. Given an effective temperature of  $T_{\mathrm{eff}\,1}$ and a gravity of ${\mathrm{log}}g$ for the primary, we obtain its age $t$, its mass $M_1$ and its luminosity $L_1$ from the evolutionary grids of the binary population synthesis (BPS) of \cite{Han2002}.
%isochrone evolutionary grids of the Binary Star Evolution (BSE) code \citep{BSE2013}. 
Given a mass ratio $q=M_2/M_1$, we obtain the mass of the secondary star $M_2$. Using the BPS model grid of \cite{Han2002} again, we obtain the luminosity $L_2$ and the associated $T_{\mathrm{eff}\,2}$ of the secondary star of age $t$.
Based upon the atmospheric parameters of the primary star $M_1$ and the secondary star $M_2$ with age $t$, we obtain the model spectra for the primary and the secondary from ATLAS9, respectively. 
%Also, we set the range of q from 0.4 to 1 in {\bf Fig.3} for clarity that the possibility of detected binary is zero when q is smaller than 0.4 where the luminosity of the secondary is too low compared to the primary's to be detected. 
We shift the secondary's spectrum with a radial velocity $\Delta RV$ and combine the spectrum with that of the primary.  The minimal radial velocity difference is set to be $50\,{\rm km/s}$ since the minimal detected $\Delta RV$ is around $50\,{\rm km/s}$ from LAMOST-MRS. The largest $\Delta RV$ added on the spectrum is $225\,{\rm km/s}$ because it is the maximum that \cite{Lichunqian2021} detected from observations. Beyond this range, the possibility of detecting the binary is set to zero.
The combined spectra are then injected with Gaussian noise with a given signal-to-noise (SNR) ratio. The ratio is larger than 10 to be consistent with the observational spectra. If the ratio is larger than 100, there is no significant influence on the detected efficiency of the spectra.

We constructed binary synthetic spectra to find a way to derive mass ratios of observed SB2s from LAMOST-MRS data.  To make things simple, we set primary stars' gravity to be ${\mathrm{log}}g=4.5$, and adopt $T_{\mathrm{eff}\,1}=8\,000 {\,\rm K}$ for primary stars of spectral type A, 6\,400\,K for type F, and 5\,600\,K for type G, respectively. We chose such temperature values as they can be found from the BPS model grids of \cite{Han2002}, but still be typical for stars of spectral type A, F, and G. 
For each spectral type of the primary, we adopt various mass ratios $q$ from 0.4 to 1.0. $q$ is chosen in such a way that $L_2/L_1$ changes by a constant step size, and $q=0.4$ corresponds to $L_2/L1=0.1$, which is the lower limit for SB2 detection. For binaries with given mass ratios $q$ and given spectral types, we 
adopt various $\Delta RV$ and SNR
 (see Table~\ref{tab:para}) to construct their synthetic spectra.

\begin{deluxetable*}{ccrrr}
\tablenum{2}
\tablecaption{Parameters for our synthetic spectra grid\label{tab:para}}
\tablewidth{0pt}
\tablehead{
\colhead{Parameter} &Value & 
}
\startdata
Wavelength  ({\AA}) & 4\,900- 5\,300, 6\,300 - 6\,900 \\
Temperature (K) & 4\,000, 4\,100, 4\,200, ... , 8\,000  \\
SNR &  10, 20, 50, 100\\
$\Delta RV$ (km/s) & 50, 60, 70, 80, 90, 100, 125, 150, 175, 200, 225\\
$q$ & 0.40, 0.50, 0.63, 0.79, 1.00\\
\enddata
\tablecomments{We have constructed a synthetic spectra grid from $ALTAS$9 with a resolution of 7\,500. The 1st, 2nd, and 3rd rows show the wavelength range, the effective temperature, and the SNR, respectively. The 4th and 5th rows are for binaries, in which the 4th row is the RV differences of the two components, and the 5th row is for the mass ratio.}
\end{deluxetable*}

\begin{figure}
\plotone{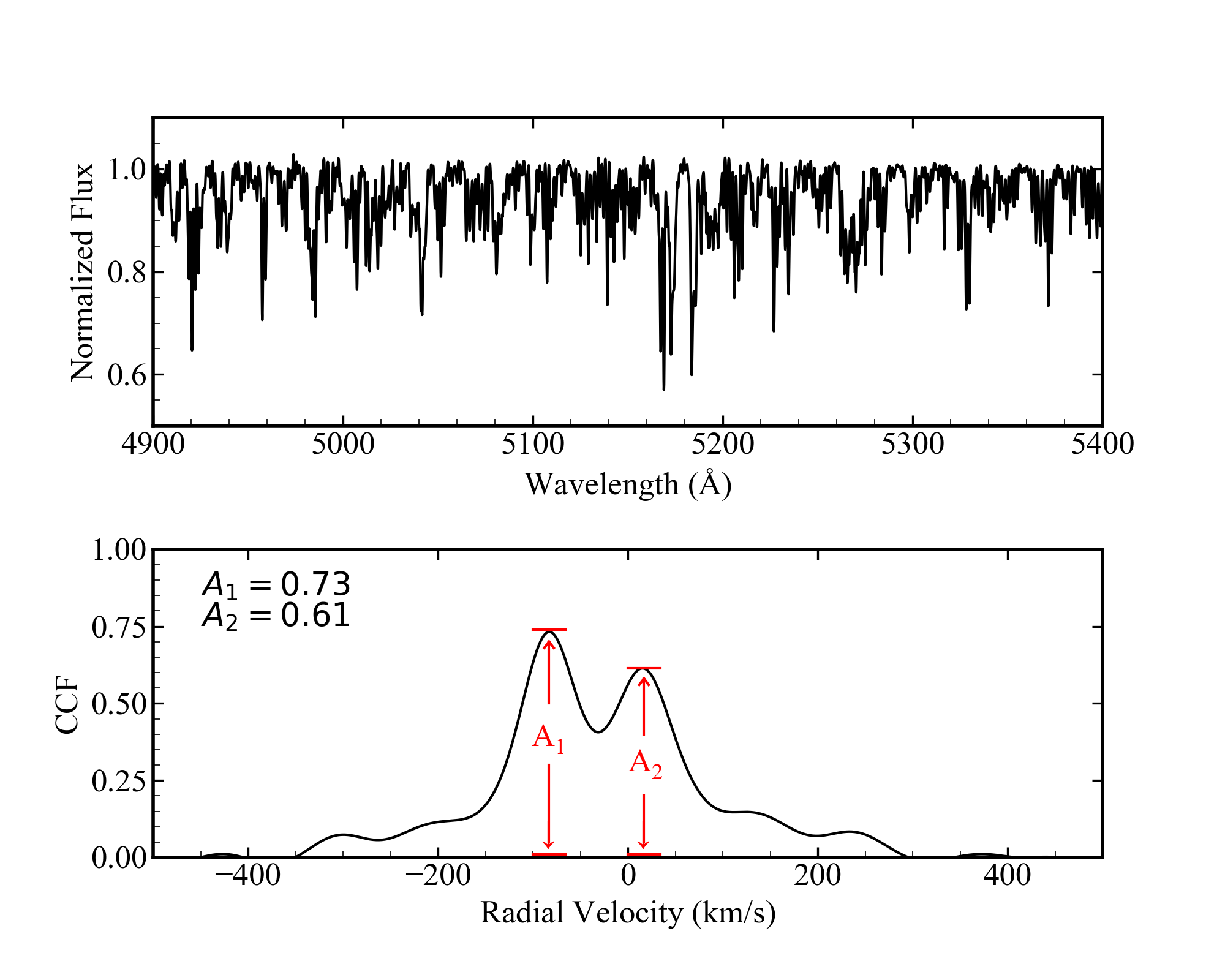}
\caption{An example for the CCF of a binary system. In the upper panel, we show a spectrum of a binary system with primary's effective temperature of $T_{\mathrm{eff\,1}}=6\,400$\,K and a mass ratio of $q=0.79$. The spectrum has a resolution of 7\,500 with a wavelength range from 4\,900\,\AA\ to 5\,400\,\AA. The lower panel shows the CCF of the spectrum, and the radial velocity ranges from $-500$ to $500$\,$\mathrm{km/s}$. There are two peaks of $A_1=0.73$ and $A_2=0.61$, and the velocities corresponding to the two peaks are $RV_1=-83$\,$\mathrm{km/s}$ and $RV_2=17$\,$\mathrm{km/s}$, respectively.
\label{fig:ccf}
}
\end{figure}

\begin{figure}[h]
\centering
\includegraphics[scale=0.7]{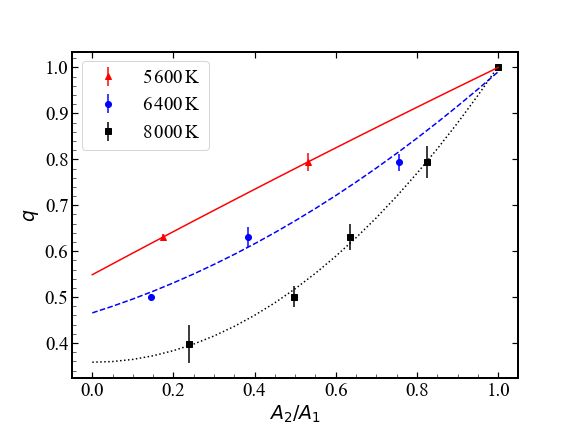}
\caption{The relation between the peak amplitude ratio (PAR), $A_2/A_1$, and the mass ratio,$q$. The red triangles and the solid line show the relation of 5\,600\,K, the blue triangles and the blue dashed line are the results of 6\,400\,K, and the black squares and the black dotted line are the results of 8\,000\,K.}
\label{fig:fits}
\end{figure}

We employ the code from \cite{Lichunqian2021} to calculate a CCF from the constructed synthetic spectra of binaries. Figure~\ref{fig:ccf} shows an example of CCF calculation. We see there are two peaks for the CCF, with the large peak for the primary star ($A_1$) and the small peak for the secondary star ($A_2$).
In the upper panel, we show a template binary spectrum with an effective temperature of $T_{\mathrm{eff\,1}}=6\,400$\,K and a mass ratio of $q=0.79$. The spectrum has a resolution of 7\,500 with a wavelength range from 4\,900\,\AA\ to 5\,400\,\AA. We show the CCF of the spectrum in the bottom panel. There are two peaks of $A_1=0.73$ and $A_2=0.61$. The radial velocity ranges from $-500$ to $500$\,$\mathrm{km/s}$. We identify the positions of $A_1$ and $A_2$ as the radial velocities of the primary and the secondary ($V_1=-83$\,$\mathrm{km/s}$ and $V_1=17$\,$\mathrm{km/s}$
The peak amplitude ratio (PAR), $A_2/A_1$, where $A_1$ and $A_2$ are the peak amplitudes for the primary and the secondary in the CCF, mainly depends on the mass ratio $q$, but also on the temperature $T_{{\rm eff}\,1}$ of the primary and the radial velocity difference $\Delta RV$.
By calculating the CCFs for all the synthetic spectra constructed for binaries, we obtained the relation between $A_2/A_1$ and $q$ for different spectral types of primaries in Figure~\ref{fig:fits}.  The error bars are the scatter resulting from various $\Delta RV$ adopted. We fitted the relation between $A_2/A_1$ and mass ratio $q$ using a quadratic function, shown in Equation~\ref{eq:1} to Equation~\ref{eq:3}. With these equations, we can convert the PAR of observational SB2s to their mass ratio.

\begin{equation}
q=0.65\times (\frac{A_2}{A_1})^2 + 0.36 \ (7\,200\,\mathrm{K}\leq {\mathrm{T_{eff}}} \leq 8\,400\,\mathrm{K})
\label{eq:1}
\end{equation} 

\begin{equation}
q=0.25\times (\frac{A_2}{A_1})^2 + 0.28\times \frac{A_2}{A_1} + 0.47 \ (6\,000\,\mathrm{K}\leq {\mathrm{T_{eff}}} \leq 7\,200\,\mathrm{K})
\label{eq:2}
\end{equation}

\begin{equation}
q=-0.02\times (\frac{A_2}{A_1})^2 + 0.47\times \frac{A_2}{A_1} + 0.55 \ (5\,200\,\mathrm{K} \leq {\mathrm{T_{eff}}} \leq 6\,000\,\mathrm{K})
\label{eq:3}
\end{equation} 

\subsection{The SB2 detection efficiency}
\label{sec: efficiency}

As pointed out by \cite{Lichunqian2021}, the radial velocity differences $\Delta RV$ between the two components of binaries are the key factor for SB2 candidate identification. The identification also depends on mass ratio $q$, as shown in the previous section (Section~\ref{sec:PAR}). A binary with a large $q$ can be more easily detected. To understand how the CCF method works and how the bias of PAR derived $q$ distribution can be corrected, we have carried out an investigation on the SB2 detection efficiency, and its dependence on both $\Delta RV$ and $q$.

We have constructed a very dense grid\footnote{The grid is somewhat unnecessarily dense for the current purpose} of synthetic spectra for binaries with various $q$ and $\Delta RV$ to investigate the SB2 detection efficiency.
The effective temperature of the primaries $T_{{\rm eff}\,1}$ is taken to be 4\,000\,K to 8\,000\,K, with a step size of 100\,K, The radial velocity difference $\Delta RV$ is taken from 50 to 225\,${\rm km/s}$ with a step size of 5\,${\rm km/s}$. For a binary with a given set of parameters of $T_{{\rm eff}\,1}$, $\Delta RV$ and $q$, we generate 10 spectra with a signal-to-noise ratio of SNR=20\footnote{We have also tested other SNRs from 10 to 100, and the results do not change.}, which is a typical value for our LAMOST-MRS data. We then apply the CCF method to the spectra to see we identify them as SB2s or single stars. We define the detection efficiency as the fraction of synthetic spectra that are identified as SB2s with the CCF method.
If $\Delta RV$ is below 50\,${\rm km/s}$ or over 225\,${\rm km/s}$, the efficiency is set to zero. For $q<0.4$, the secondary is far too dimmer than the primary that the efficiency is zero.

\begin{figure}[h]
\plotone{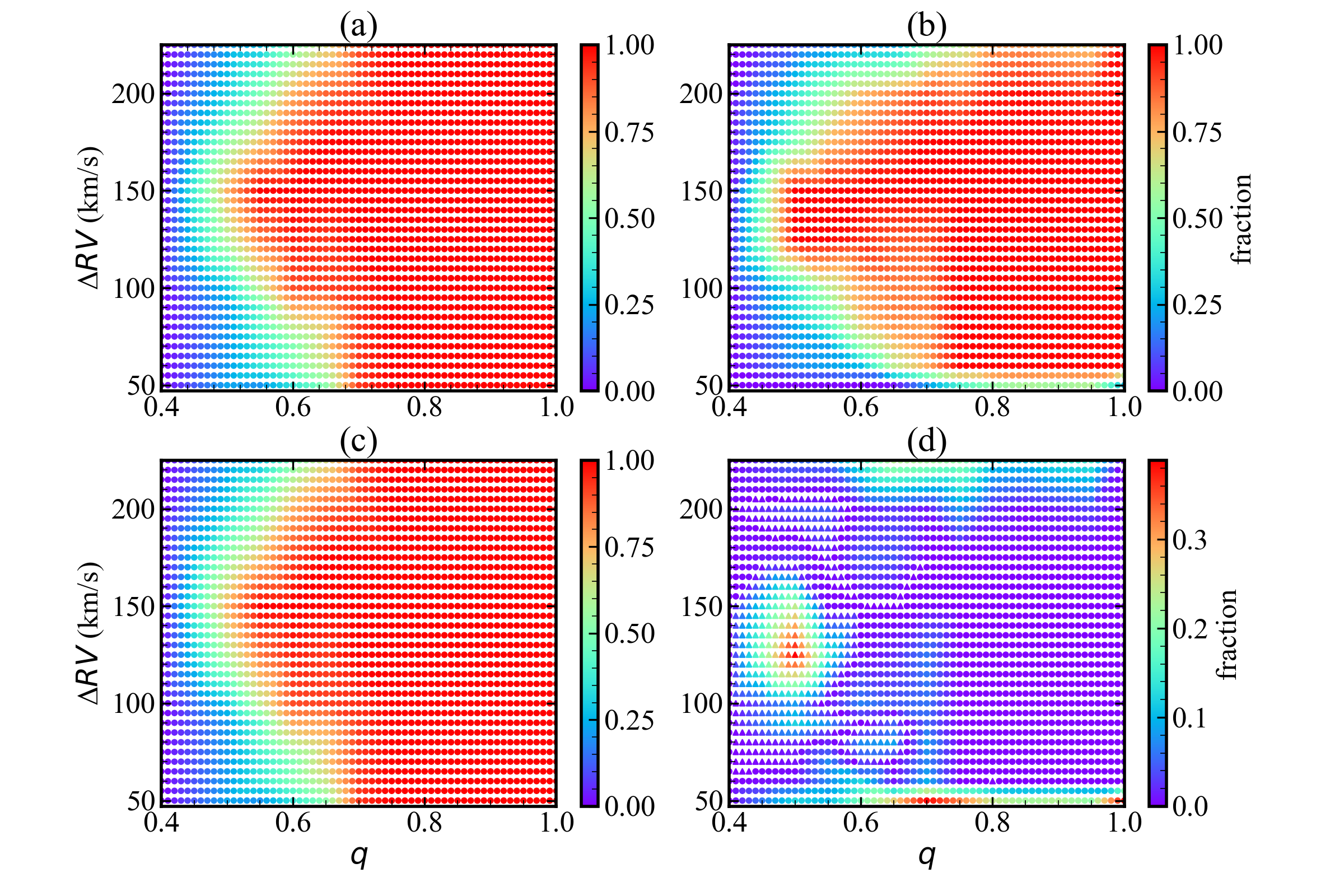}
	\caption{The dependence of detection efficiency on mass ratio $q$ and RV difference $\Delta RV$ of SB2s.  Panel (a) is for the temperature of the primary star between 6\,000\,K and 8\,000\,K,  panel (b) is for the temperature between 4\,000\,K and 6\,000\,K, while panel (c) is for all the temperatures from 4\,000 to 8\,000\,K. Panel (d) shows the residuals of panel (a) minus panel (b), in which circles are for positive values and triangles for negative values.
	\label{fig:colorbar_q_rv}}
\end{figure}

Figure~\ref{fig:colorbar_q_rv} shows the detection efficiency of SB2s for various $\Delta RV$. Panel (a) 
is for the primary's temperature between 6\,000\,K and 8\,000\,K, panel (b) for the primary's temperature between 4\,000\,K and 6\,000\,K， panel (c) for all the temperatures from 4\,000\,K to 8\,000\,K. Panel (d) shows the residuals of panel (a) minus panel (b). From panels (a), (b) and (d), we see that the SB2 detection efficiency are nearly the same for different temperatures. 
Therefore, we use panel (c), which is for all the temperatures, for the detection efficiency analysis.
From panel (c), we see that the efficiency is high (close to 1) for a large $q$ and low for a small $q$.
The low detection efficiency for a small $q$ is due to that the secondary's contribution to the synthetic spectra becomes insignificant when $q$ decreases.
We also see that SB2s can most probably be detected when the RV difference is $\sim 150\,{\rm km/s}$ for $q>0.7$ (see the right part of the panel (c)), and we explain this below.
A spectral line in a component star manifests itself as two lines in the combined spectra due to an adopted RV difference between the two-component stars. 
A large $\Delta RV$ makes the separation between the two lines bigger, and the synthetic spectra are more obvious to be an SB2. Therefore large $\Delta RV$ are favored for the detection of SB2s. However, this is not always true. $\Delta RV$ of $\sim 150\,{\rm km/s}$ corresponds to a separation of lines of $\sim 3$\,{\AA}.  Our template is the Sun spectrum of high-resolution \citep{Kurucz1993} which \cite{Lichunqian2021} used, and we have many absorption lines of iron in the Sun spectrum with slight wavelength difference. The lines with a large RV difference are easy to be identified as a new line of the single star. As a result, the efficiency decreases with increasing RV differences if RV differences are larger than $\sim 150\,{\rm km/s}$.

\begin{figure}[h]
\centering
\includegraphics[scale=0.7]{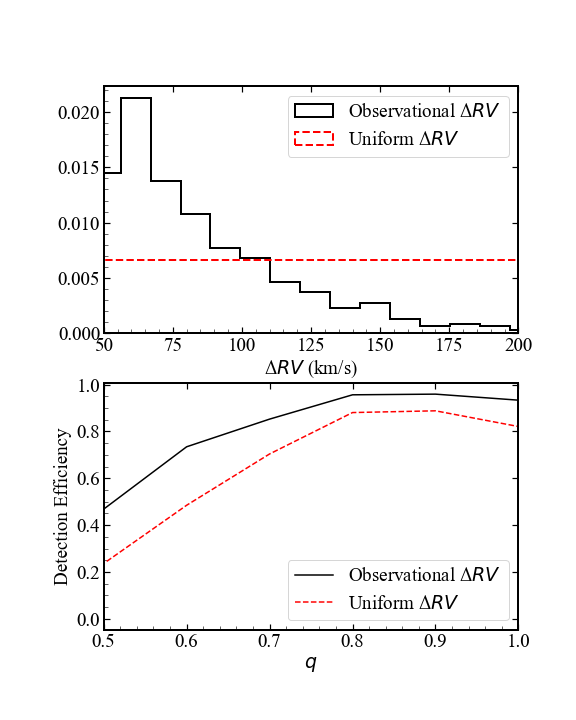}
	\caption{In the upper panel, we show the distribution of radial velocity differences $\Delta RV$ between components of the observed SB2s as a black solid line, and a uniformly distributed $\Delta RV$ as a red dashed line.  In the lower panel, we show the dependence of detection efficiency on mass ratio $q$ with observed $\Delta RV$  distribution (black solid line) and the uniformly distributed $\Delta RV$ (red dashed line). 
	\label{fig:RVs}}
\end{figure}

To have a better look at how the detection efficiency depends on $q$, we plotted Figure~\ref{fig:RVs}. If we assume that the SB2s have $\Delta RV$ uniformly distributed between 50 and 225\,${\rm km/s}$, we sum up the efficiencies of all the $\Delta RV$s for every $q$ in panel (c), divide the sum by the number of data points, and we then obtain the efficiency dependence on $q$. We show the dependence curve by the red dashed line in the lower panel. However, $\Delta RV$ of our observed SB2s is not uniformly distributed between 50 and 225\,${\rm km/s}$, and the distribution of observed $\Delta RV$ peaks at $\sim 60\,{\rm km/s}$ (see the upper panel of Figure~\ref{fig:RVs}). 
We again average the efficiencies of various $\Delta RV$s for every $q$ in panel (c), but with a weight of observed $\Delta RV$ distribution. The result is shown as the black solid line in the lower panel.
We see that the detection efficiencies with the observational $\Delta RV$ distribution are higher than the one with a uniform $\Delta RV$ distribution for $q$ from 0.4 to 1.0.

\section{Results}
 \label{sec:results}
\begin{figure}[h]
\centering
\includegraphics[scale=0.7]{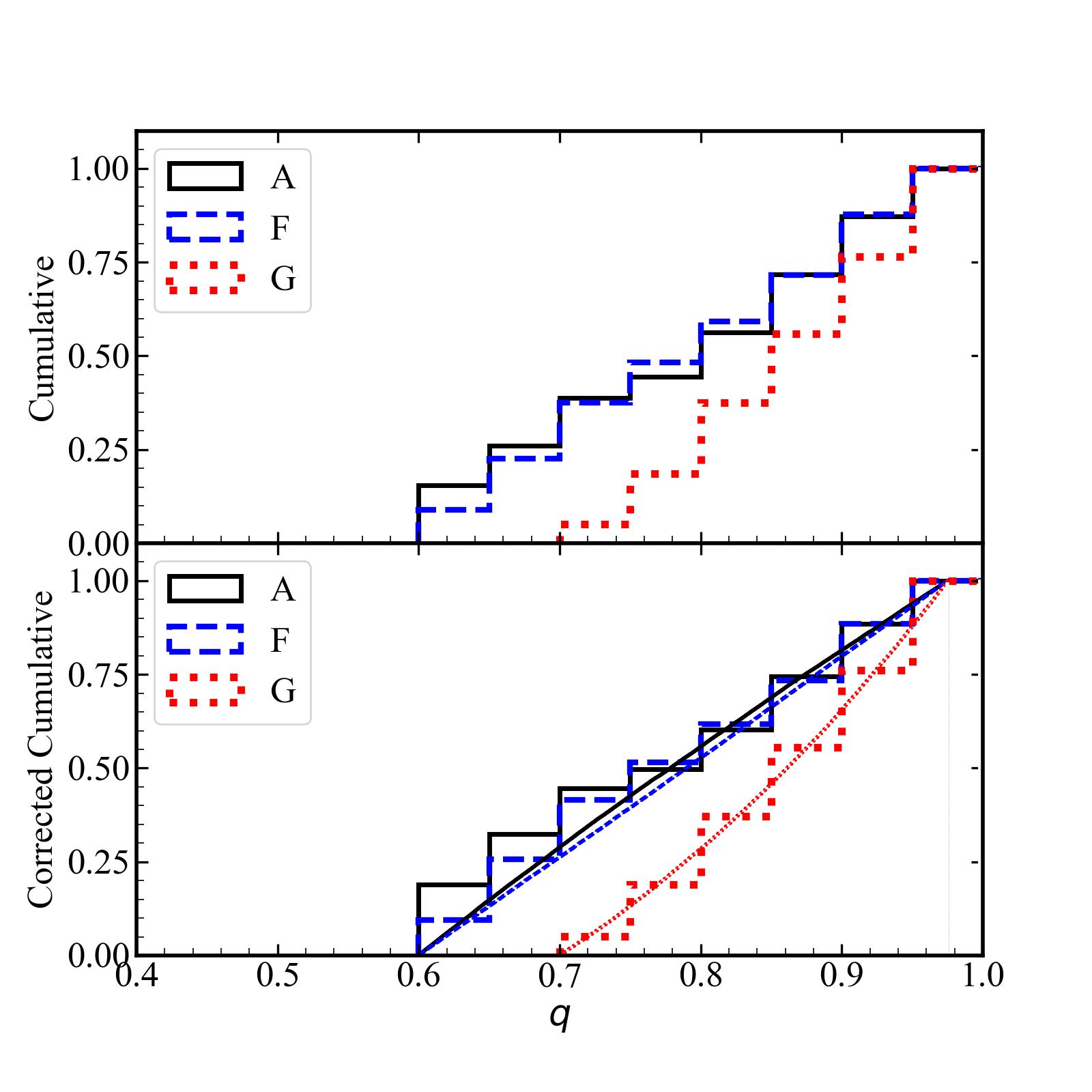}
\caption{The cumulative mass ratio distribution of A, F and, G type binaries. The distributions derived from the CCFs are shown in the upper panel, and the bias-corrected distributions and the fits are shown in the lower panel. In the figure, the black solid lines are for A-type binaries, the blue dashed lines for F type binaries, and the red dotted lines for the G type binaries.The power index for the corrected mass ratio distribution for A-type binaries is $-0.42\pm0.27$, for F-type binaries $0.03\pm0.12$, for G-type binaries $2.12\pm0.19$.}
\label{fig:q-distr}
\end{figure}

Employing the code from \cite{Lichunqian2021}, we have calculated the CCFs from the spectra of observed SB2s of LAMOST-MRS, and obtained their peak amplitude ratios (PARs). 
We then convert the PARs to mass ratios with Equation~\ref{eq:1} to Equation~\ref{eq:3}.
The upper panel of Figure~\ref{fig:q-distr} is the result for the observed SB2s of LAMOST-MRS, which shows the derived mass ratio distributions for SB2s of spectral type-A, -F, and -G.  

However, the mass ratio distribution is biased due to that a binary with a large $q$ and a $\Delta RV$ around 150\,${\rm km/s}$ is more easily detected from the CCF. In other words, the detection efficiency depends on $q$ and $\Delta RV$. We have obtained the dependence of the detection efficiency on $q$ and $\Delta RV$ in Section~\ref{sec: efficiency} for binaries (Figure~\ref{fig:colorbar_q_rv}).
Using the efficiency in panel (c) of Figure~\ref{fig:colorbar_q_rv},  we corrected the mass ratio distribution of the upper panel of Figure~\ref{fig:q-distr}. The bias-corrected distributions are shown in the lower panel of Figure~\ref{fig:q-distr} for binaries of spectral type A, F, and G.
As seen from Figure~\ref{fig:colorbar_q_rv}, the detection efficiency is low for $q \sim 0.5$ and $\Delta RV<50\,\mathrm{km/s}$, which means that the corrected distribution for small mass ratios is not very reliable. We therefore only use the bias-corrected distribution of $q\geq 0.6$ in our analysis in this paper.

We assume that the bias-corrected mass ratio distribution follows a power-law function of $q^{\gamma}$. By using a power-law fitting of the corrected mass ratio distribution ${\rm d }N/{\rm d}q\propto q^{\gamma}$, the $\gamma$ of mass ratio distributions of SB2 candidate systems are estimated to be $-0.42\pm0.27$, $0.03\pm0.12$ and $2.12\pm0.19$ for A, F and G type stars, respectively. The power law index of mass-ratio distribution increases towards lower mass, suggesting that the less massive star tends to harbor a companion star with a comparable mass. 
%In the bottom panel of Figure~\ref{fig:bias}, the corrected mass ratio distribution is calculated by using the observational mass ratio distribution (upper panel of Figure~\ref{fig:bias}) and the functions of the detection efficiency (middle panel of Figure~\ref{fig:bias}). Under the aforementioned methods, the $\gamma$ of mass-ratio distributions of SB2 candidate systems are estimated as to be $-3.89\pm0.06$, $-1.85\pm0.07$ and $1.00\pm0.13$ for A, F and G type stars, respectively. The power law index of mass-ratio distribution increases towards lower mass, suggesting that the less massive star tends to harbor a companion star with q comparable mass. 

\section{Discussion}
\label{sec:discussion}

\subsection{Justification for ${\mathrm{log}}g=4.5$ adopted in our model}
\begin{figure}[h]
\centering
\includegraphics[scale=0.7]{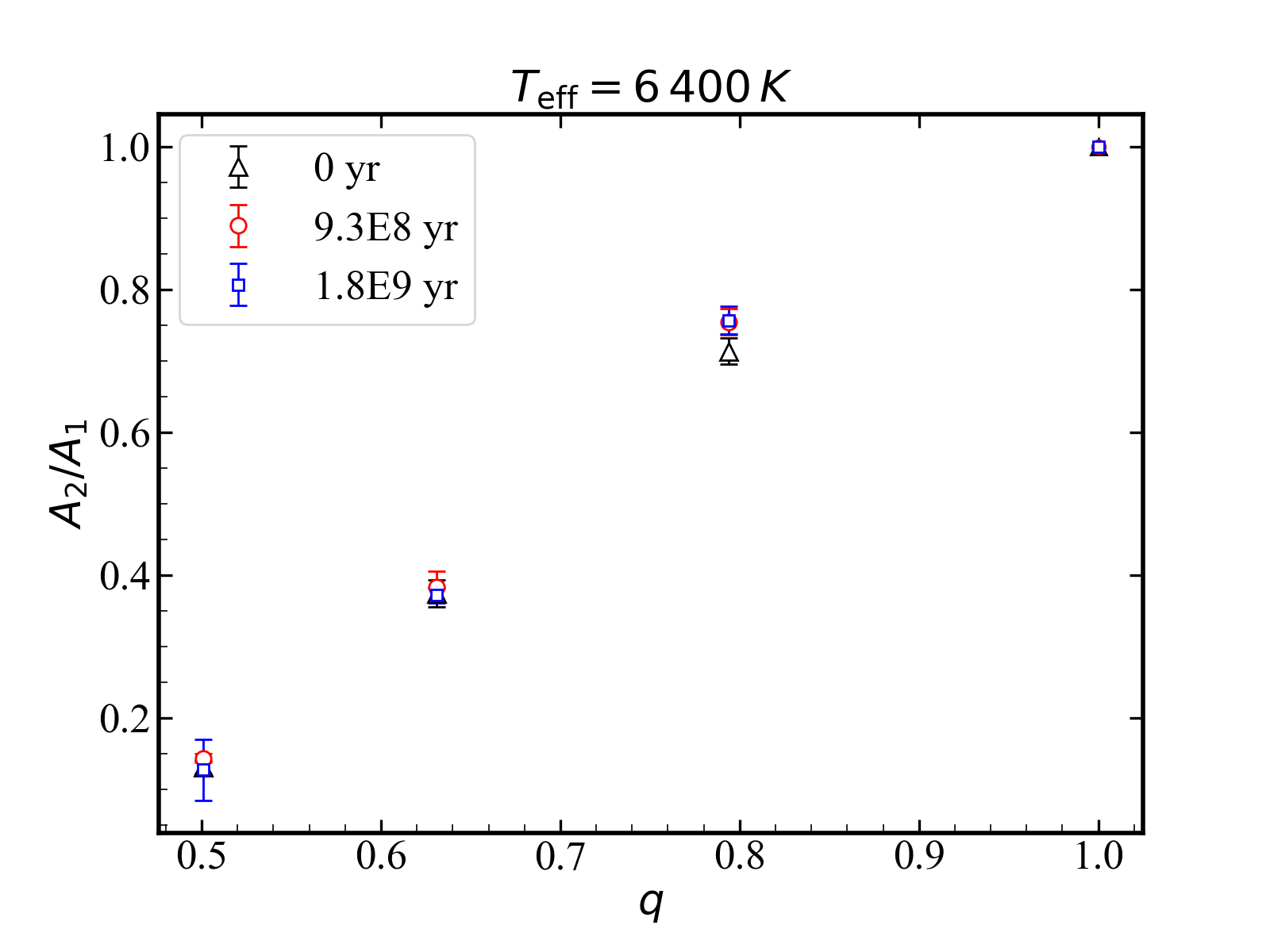}
\caption{The relation between PAR and the mass ratio. Black triangles are for binaries at zero age main sequence (ZAMS) ($t=0\,{\rm yr}$), red circles for binaries at the middle age of the MS (MAMS) ($t=9.3\times10^{8}$\,{\rm yr}) and blue squares for binaries at the terminal age MS (TAMS) ($t=1.8\times10^{9}$\,{\rm yr}).}
\label{fig:6400}
\end{figure}

We have constructed binary synthetic spectra in order to find a way to derive mass ratios of observed SB2s from LAMOST-MRS data.  To make things simple, we have assumed ${\mathrm{log}}g=4.5$ for the primary. We show below that such an assumption is suitable for our purpose. For a binary system of spectral type F, the effective temperature of the primary 
$T_{\mathrm{eff}\,1}$ is set to be 6\,400K. We set $t=$ 0\,yr, $9.3\times 10^8$\,yr, and $1.8 \times 10^9$\,yr, corresponding to the zero-age MS (ZAMS), the middle age MS (MAMS), and terminal age MS (TAMS) of the primary star, respectively. The gravity, ${\mathrm{log}}g$, of the primary decrease from ZAMS to TAMS, but its value is not far from 4.5. 
We then constructed synthetic spectra of the binary with various mass ratios for three ages $t$.
Employing the code from \cite{Lichunqian2021}, we calculated CCFs from the constructed synthetic spectra, and Figure~\ref{fig:6400} shows the result for the relation between mass ratio $q$ and PAR $A_2/A1$. We see very little difference among ZAMS, MAMS, and TAMS primaries. Adopting ${\mathrm{log}}g=4.5$ would not noticeably affect our result.

\subsection{The dependence of SB2 detection on orbital periods}
\label{sec:periods}
\begin{figure}[h]
\centering
\includegraphics[scale=0.7]{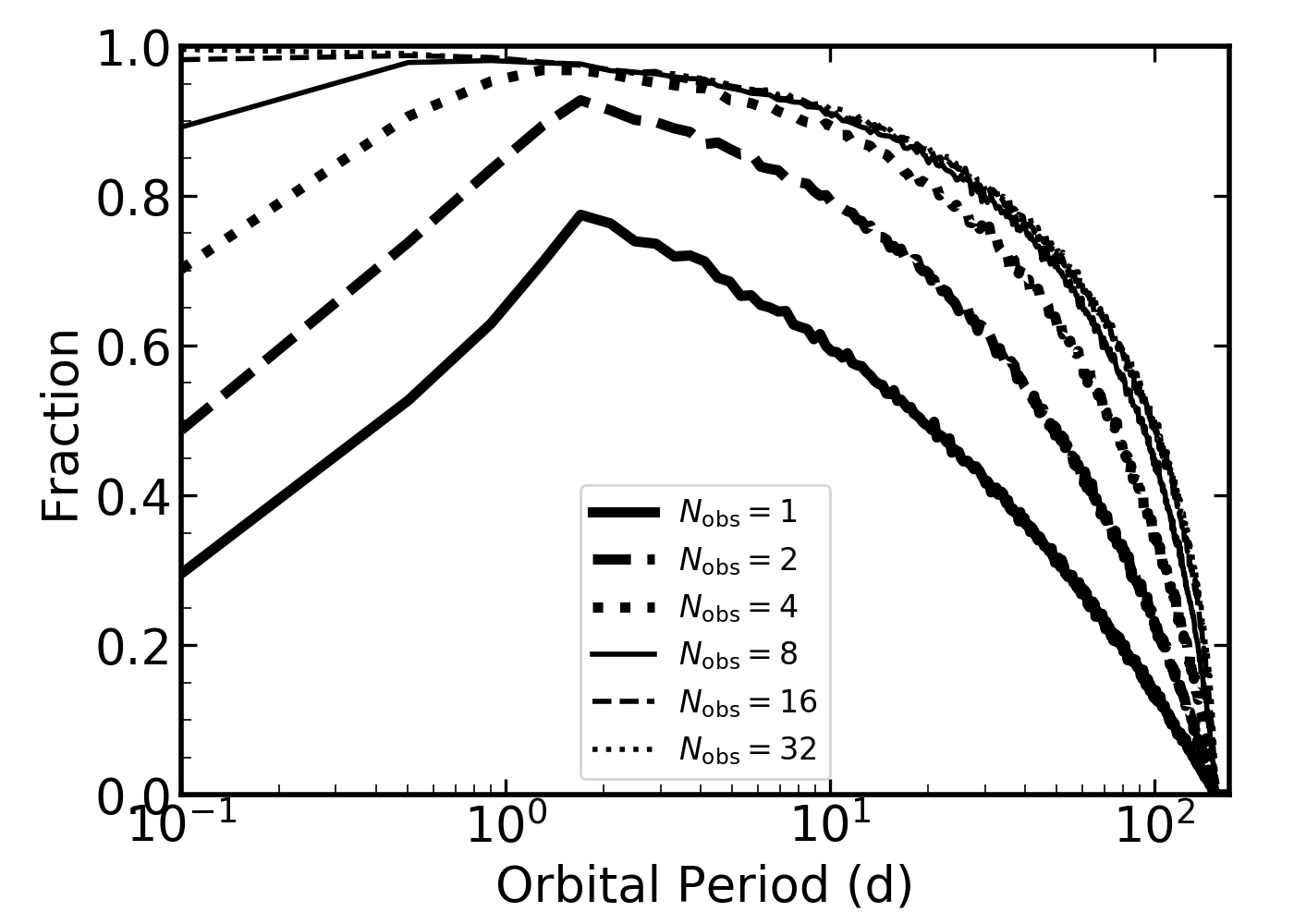}
	\caption{The fraction of detected SB2s versus orbital period for a given number of observational epochs. The thick solid, dashed, and dotted lines are for 1, 2, and 4 observational epochs, while
	the thin solid, dashed, and dotted lines are for 8, 16, and 32 observational epochs.
	\label{fig:ratio}}
\end{figure}

In our analysis for mass ratio distributions, we have corrected the bias resulting from the dependence of SB2 detection efficiency on $q$ and $\Delta RV$. The detection also depends on the orbital period and the number of observation epochs for an SB2, and such a bias needs to be discussed. We show the detection fraction of SB2s versus orbital period in Figure~\ref{fig:ratio}.

In the figure, we assume the total mass of the primary and the secondary to be $2\,M_{\odot}$, the orbital eccentricity is zero, the mass ratio is one, the inclination ranges from 0 to $2\pi$ (uniformly distributed in solid angles), and the observational epoch is uniformly distributed at orbital phase between 0 and $2\pi$.  For a binary with a given orbital period, we can then calculate the radial velocity difference $\Delta V$ of the observational epoch. We did such a calculation for 10\,000 binaries with a given number of observational epochs and a given orbital period. We define the detection fraction as the ratio of the number of binaries with $\Delta RV$ between 50 and 225\,${\rm km/s}$ over the number of the total binaries (i.e. 10\,000).
We see that the CCF method we adopted is able to detect SB2s with orbital periods less than 155\,d, and most of the detected SB2s are with orbital periods less than 50\,d.

\subsection{Completeness}
\begin{figure}[h]
\centering
\includegraphics[scale=0.6]{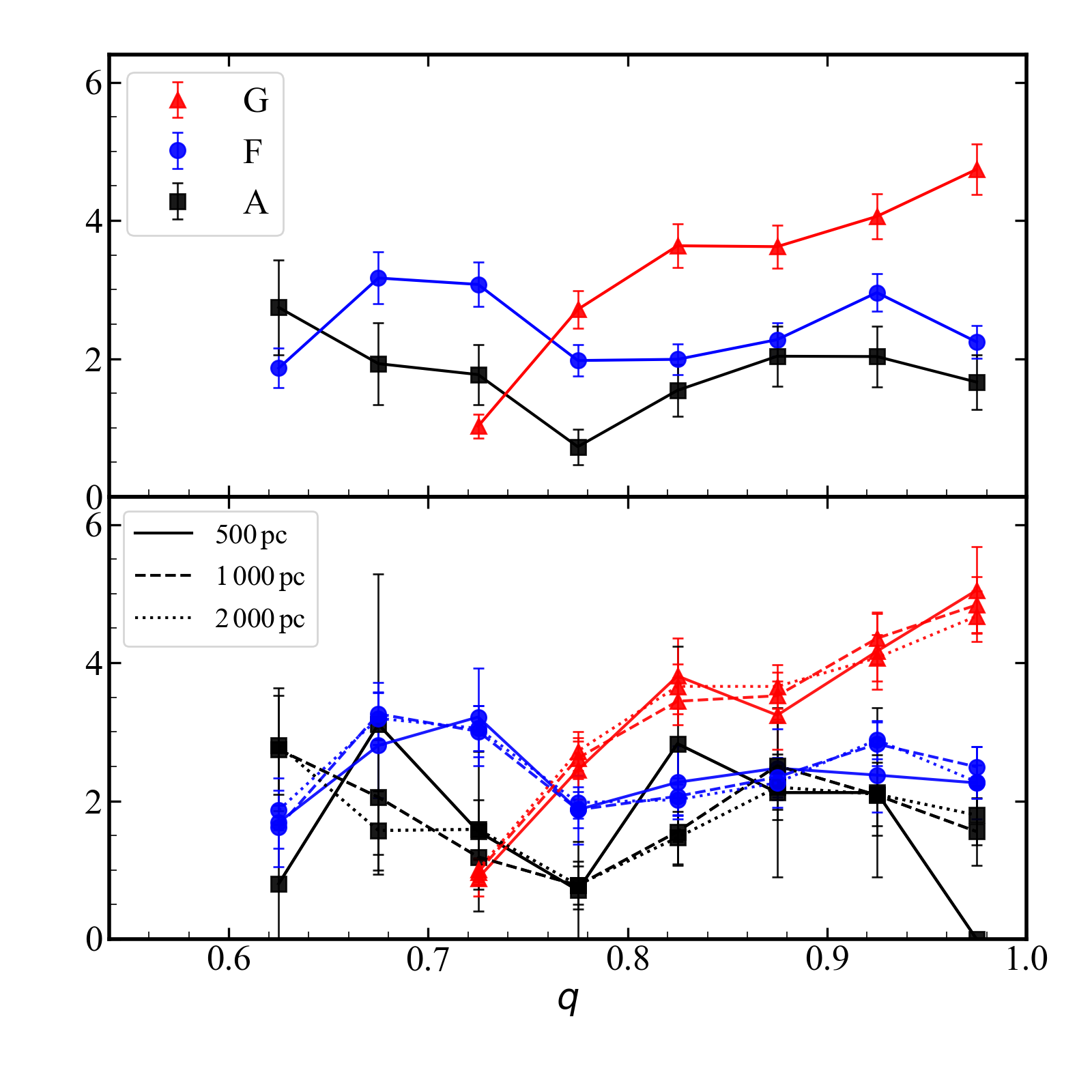}
\caption{Mass ratio distribution for binaries with different distances to the Sun. In the upper panel, we show the distributions of A, F, and G type binaries (black squares, blue circles, and red triangles) of our sample. The bottom panel shows similar distributions, but for binaries within 500, 1\,000, and 2\,000\,pc of the Sun (solid, dashed, and dotted lines). Note that error bars in the lower panel overlap for a given $q$, and bigger error bars are for small sample sizes or small distances to the Sun.}
\label{fig:density}
\end{figure}

The LAMOST-MRS is a magnitude-limited survey in which the detection limitation is around 14.5 apparent magnitude in G band \citep{Liu2019}. For binaries with the same effective temperature, a binary with a large mass ratio is brighter than the one with a small mass ratio. It may cause a bias that the $\gamma$ of the observed mass ratio distribution could be higher than the true value. The distance of our candidates is below 3\,000\,pc. For considering the completeness of our result, we make three volume-limited samples with a distance to the Sun of 500, 1\,000, and 2\,000\,pc, respectively. Using the number of identified SB2 stars in Table \ref{tab:1} for different distances to the Sun and the same procedure described in Section~\ref{sec:results}, we plot the mass ratio distribution for binaries of spectral type A, F, and G in Figure~\ref{fig:density}. Note that the distributions are not cumulative ones as in Figure~\ref{fig:q-distr}.   
We see no significant differences between the distributions for different distances (see the bottom panel in Figure~\ref{fig:density}). Therefore, we conclude that our results are affected little by the bias of the magnitude-limited sample.

\subsection{Binary fractions of type-A, -F, and -G stars}
\begin{figure}[h]
\centering
\includegraphics[scale=0.7]{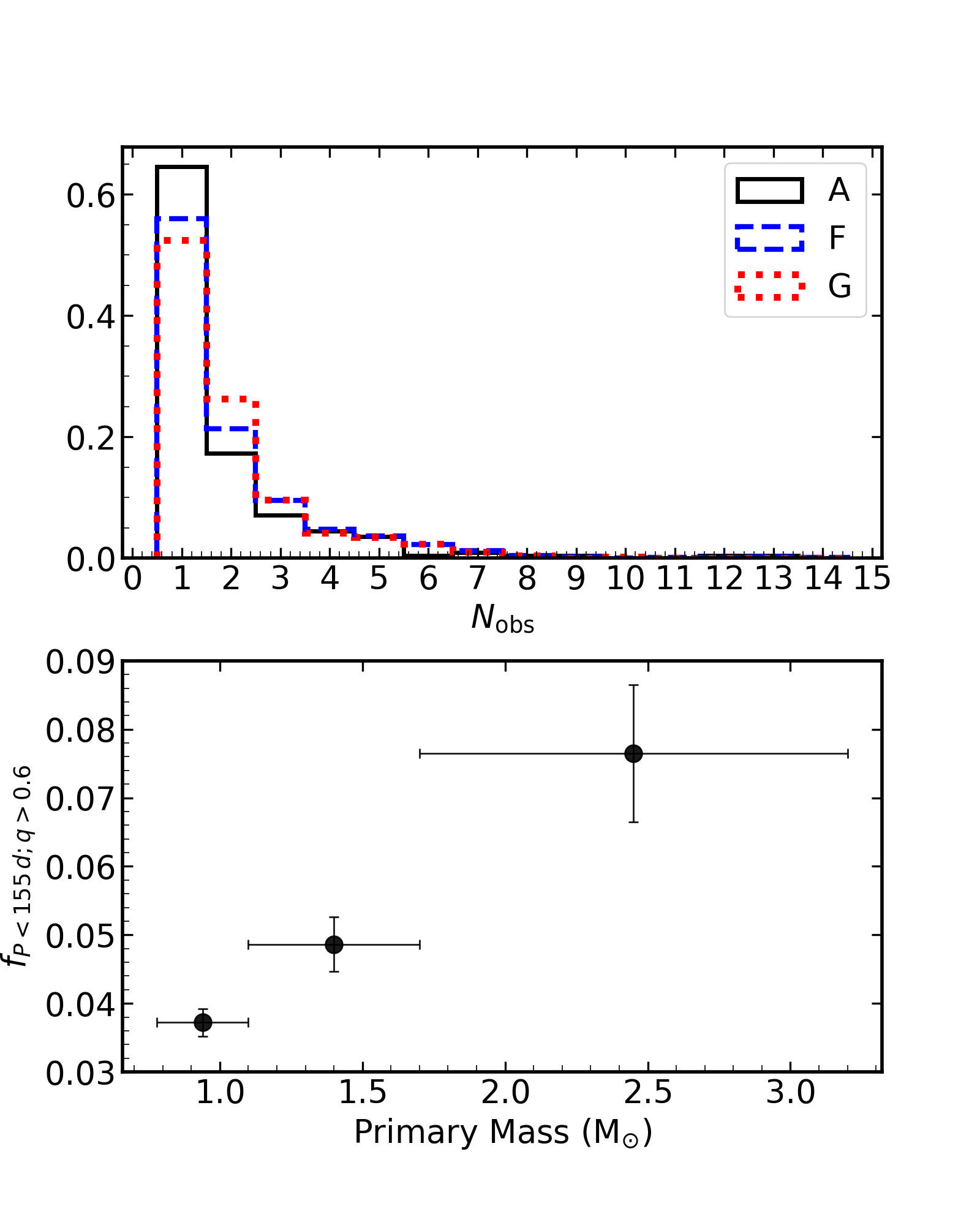}
	\caption{The upper panel shows the distributions of the number of observational epochs of A, F, and G type SB2s in our sample (black solid, blue dashed, and red dotted lines, respectively).  The lower panel shows the binary fractions with an orbital period of less than 155 days and a mass ratio between 0.6 to 1. The close binary fractions for A, F and G binaries are $7.6\pm 0.5 \%$, $4.9\pm 0.2 \%$ and $3.7 \pm 0.1 \%$, respectively.
	\label{fig:nobs}}
\end{figure}

In Section~\ref{sec:periods}, we have discussed how the detection efficiency depends on the orbital period and the number of observational epochs of an SB2. 
The upper panel of Figure~\ref{fig:nobs} shows the distribution of observational epoch numbers of A, F, G type SB2s in our sample. 
Our SB2s have 1 to 15 observations.
More than 55\% SB2s have only one observation, and $\sim 96\%$ SB2s have been observed less than 5 times. With the detection fraction shown in Figure~\ref{fig:ratio} and the detection efficiency shown in Figure~\ref{fig:colorbar_q_rv}, we convert the detected number of SB2s to the intrinsic one, and obtain the binary fraction by dividing the intrinsic number of binaries by the total number of stars observed.

The lower panel of Figure~\ref{fig:nobs} shows the fraction of close binaries with an orbital period less than 155\,d and a mass ratio between 0.6 and 1 for spectral type A, F, and G. The close binary fractions for A, F and G binaries are $7.6\pm 0.5 \%$, $4.9\pm 0.2 \%$ and $3.7 \pm 0.1 \%$, respectively. The close binary fractions we derived are consistent with that of \cite{Moe2017}, in which the multiplicity of solar-type primaries $f_{q>0.3}=0.017\pm0.007$ with $\mathrm{log}(P/\mathrm{ d})=0.5\pm0.5$ and $f_{q>0.3}=0.020\pm0.007$ with $\mathrm{log}(P/\mathrm{ d})=1.5\pm0.5$. Also, in \cite{Moe2017}, they showed that binary fraction increase with stellar mass,  and estimated $f_{P<20\,{\mathrm{d}}}\propto M_1^{0.7}$, where A-dwarfs with $M_1= 2.45\pm0.75$ $M_{\odot}$ would have $\sim$ 1.96 times the close binary fraction of G-dwarfs with $M_1 = 0.94 \pm 0.16 M_{\odot}$. 

\subsection{Comparison with Previous Work}
\begin{figure}[h]
\centering
\includegraphics[scale=0.5]{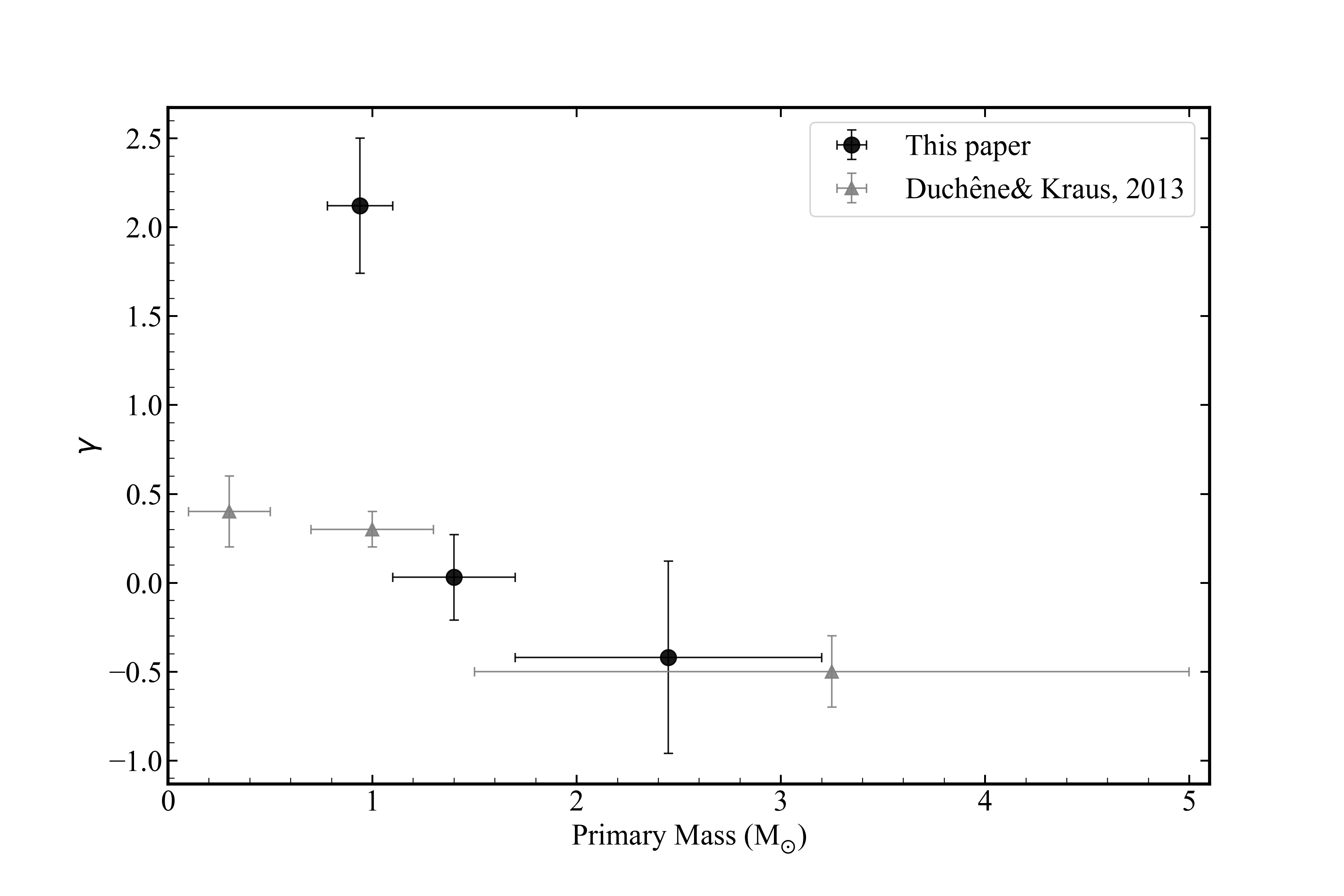}
\caption{The power-law index $\gamma$ of mass-ratio distribution as a function of the primary mass. Black circles with black error bars show our results, and grey triangles denote the results from \cite{Duchene2013} for primary stars with masses larger than 0.3 $M_{\odot}$ with mass ratios down to 0.1.}
\label{fig:comparison}
\end{figure}

%We compare our result with the conclusion derived by \citet{Duchene2013} shown in Figure~\ref{fig:comparison}. 
By using the binary samples of \citet{Raghavan2010}, \citet{Shatsky2002} and \citet{Kouwenhoven2007}, \citet{Duchene2013} found an increasing trend of the power index $\gamma$ of mass ratio distribution towards lower primary mass, shown as gray circles in Figure~\ref{fig:comparison}.
In our study, the index $\gamma$ is marked by black circles in Figure~\ref{fig:comparison}, and they display a similar trend. However, from Figure~\ref{fig:comparison}, we see that the variation of index $\gamma$ with the primary mass in our results is much larger than their results. This may be caused by two reasons. Firstly, the mass ratio distribution may depend on the orbital periods. 
In our sample of SB2 candidates, the binaries have an orbital period of less than 155 days as shown in Figure~\ref{fig:ratio}, while the orbital periods of binaries from \citet{Duchene2013} are mainly larger than 100 days.  Actually,
there is a compelling discrepancy between the close binaries and wide binaries both on the statistical distribution and binary formation \citep{Moe2017}.
Secondly, our mass ratio ranges from 0.6 to 1 while the mass ratio in \citet{Duchene2013} ranges from 0 to 1. The range of mass ratio may lead to different mass ratio distribution \citep{Moe2017}, which may indicate the variation of the mechanism for binary formation and evolution.

%Based upon the forthcoming observations carried out by the LAMOST, we expect to enlarge the current sample of SB2 stars.
%and we will derive the mass-ratio distribution for wide binary system in future work.

\subsection{Implications for the excess of type G-binaries with equal component masses}
Figure~\ref{fig:density} shows the mass ratio distributions of binaries of spectral type A, F, and G, and we see that type-G binaries tend to have comparable component masses. Such a phenomenon is also shown by the much bigger $\gamma$ we obtained for G-type binaries in Figure~\ref{fig:comparison}. Such an excess of "twin" close binaries provides a valuable constraint on star formation mechanism and is helpful for the explanation of the origin of double white dwarfs (DWDs) with $q\sim 1$, which are potential progenitors of type Ia supernovae (SNe Ia) and also gravitational wave sources for space gravitational wave detectors, e.g. LISA, TianQin, and Taiji.

There are two major channels for the formation of binaries, i.e. fragmentation and accretion. Fragmentation tends to produce binaries with very different component masses, while accretion tends to produce binaries with more equal component masses. G-type binaries are low mass binaries, and the secondaries of the pre-MS binaries (seed binaries) accrete matter from the infalling gaseous material (molecular cloud). The preference growth of the secondary is due to the fact that the infalling matter has bigger specific angular momentum than that of the pre-MS binary, and the pre-MS binary accumulates matter to reach $q\sim 1$. For a low mass binary from such a scenario, the secondary star is younger than the primary star by a few million years. For a thorough discussion of the formation of twin binaries, we refer to \cite{Bate1997,Bate1998,Bate2018,Tokovinin1992,Tokovinin2000,Tokovinin2020,Simon2009,Offner2016,Offner2022}

Recent observations of DWDs \citep{Kruckow2021} show that a large fraction of DWDs has a mass ratio of $q\sim 1$, posing a challenge to the binary evolution theory. Conventional binary population synthesis (BPS) is difficult to explain the excess of DWDs of $q\sim 1$ as most of the DWDs are expected to have mass ratios around $\sim 0.5$ (\cite{Han1998}). 
To get out of the dilemma, Li et al. (2022, in preparation) proposed a model. In their model, the primary of a binary system evolves to a giant and fills its Roche lobe, leading to a stable mass transfer from the primary to the secondary and producing a binary with a WD and an accreted secondary. The accreted secondary continues to evolve and becomes a giant, filling its Roche lobe and the mass transfer from the giant to the WD results in the formation of a common envelope (CE). The ejection of the CE results in the formation of a DWD with $q\sim 1$. 
To make the model work, they need to make the mass transfer of a giant binary stabilized. For this purpose, they adopted the adiabatic mass loss model of \cite{Ge2010,Ge2015,Ge2020}, which shows the mass transfer of a giant binary (a giant plus an MS star) can be stable if the mass ratio (primary to secondary) is larger than 1. This is contrary to what the widely adopted polytropic model \cite{Hjellming1987} has predicted that the mass transfer is dynamically unstable. On the other hand, a large mass ratio (primary to secondary) tends to make the mass transfer of a giant binary more unstable. An excess of binaries of $q\sim 1$ (secondary to primary) will make the mass transfer later on more stable, which is a great help for the formation of DWDs with $q\sim 1$.

\section{Conclusions}
\label{sec:conclusion}

In this paper, we devised a PAR approach to derive mass ratios from the peak amplitude ratios (PARs) of cross-correlation functions (CCFs) between binaries' spectra and solar spectrum. 
We then studied the mass ratio distribution for the LAMOST-MRS data with the approach.
If we fit the distribution to a power law, we obtain a power index $\gamma$
of $-0.42\pm0.27$, $0.03\pm0.12$ and $2.12\pm0.19$ for A-, F- and G-type stars, respectively. And more importantly, type-G binaries are found to be more likely in twins, which are useful for the study of the binary formation, double white dwarfs, etc.
The binary fractions for A, F and G stars are $7.6\pm 0.5 \%$, $4.9\pm 0.2 \%$ and $3.7 \pm 0.1 \%$, respectively. Note that the mass ratio distribution and binary fractions derived here are for binaries with orbital periods less than 155\,d and mass ratios larger than 0.6.

\acknowledgments
We thank our referee, Max Moe, for his constructive and insightful comments which helped to improve the paper significantly.
This work is supported by National Natural Science Foundation of China under grant numbers of Nos. 12090040/3, 12125303, 11733008, the National Key R\&D Program of China with No. 2021YFA1600401/3, the China Manned Space Project of No. CMS-CSST-2021-A10, Yunnan Fundamental Research Projects (grant NO. 202101AV070001), the joint funding of Astronomy of the National Natural Science Foundation of China and the Chinese Academy of Sciences, under Grants No. U1831125 as well as the Research Program of Frontier Sciences, CAS, Grant NO. QYZDY-SSW-SLH007. We acknowledge the computing time granted by the Yunnan Observatories for the usage of the Phoenix Supercomputing Platform.  Guoshoujing Telescope (LAMOST) is a National Major Scientific Project built by the Chinese Academy of Sciences. Funding for the project has been provided by the National Development and Reform Commission. LAMOST is operated and managed by the National Astronomical Observatories, Chinese Academy of Sciences. This work has made use of data from the European Space Agency (ESA) mission {\it Gaia} (\url{https://www.cosmos.esa.int/gaia}), processed by the {\it Gaia} Data Processing and Analysis Consortium (DPAC, \url{https://www.cosmos.esa.int/web/gaia/dpac/consortium}). Funding for the DPAC has been provided by national institutions, in particular the institutions participating in the {\it Gaia} Multilateral Agreement.

\software{LASP pipeline \citep{Wu2011, Wu2014, Luo2015}, ULySS \citep{Koleva2009}, SPECTRUM \citep{Gray1994}, laspec \citep{Zhang2020, Zhang2021}, Padova and Trieste Stellar Evolution Code \citep{Bressan2012}}

\bibliography{ms}{}

\begin{thebibliography}{}
\expandafter\ifx\csname natexlab\endcsname\relax\def\natexlab#1{#1}\fi
\providecommand{\url}[1]{\href{#1}{#1}}
\providecommand{\dodoi}[1]{doi:~\href{http://doi.org/#1}{\nolinkurl{#1}}}
\providecommand{\doeprint}[1]{\href{http://ascl.net/#1}{\nolinkurl{http://ascl.net/#1}}}
\providecommand{\doarXiv}[1]{\href{https://arxiv.org/abs/#1}{\nolinkurl{https://arxiv.org/abs/#1}}}

\bibitem[{{Abt} {et~al.}(1990){Abt}, {Gomez}, \& {Levy}}]{Abt1990}
{Abt}, H.~A., {Gomez}, A.~E., \& {Levy}, S.~G. 1990, \apjs, 74, 551,
  \dodoi{10.1086/191508}

\bibitem[{{Arenou} {et~al.}(2018){Arenou}, {Luri}, {Babusiaux}, {Fabricius},
  {Helmi}, {Muraveva}, {Robin}, {Spoto}, {Vallenari}, {Antoja},
  {Cantat-Gaudin}, {Jordi}, {Leclerc}, {Reyl{\'e}}, {Romero-G{\'o}mez}, {Shih},
  {Soria}, {Barache}, {Bossini}, {Bragaglia}, {Breddels}, {Fabrizio},
  {Lambert}, {Marrese}, {Massari}, {Moitinho}, {Robichon}, {Ruiz-Dern},
  {Sordo}, {Veljanoski}, {Eyer}, {Jasniewicz}, {Pancino}, {Soubiran}, {Spagna},
  {Tanga}, {Turon}, \& {Zurbach}}]{Arenou2018}
{Arenou}, F., {Luri}, X., {Babusiaux}, C., {et~al.} 2018, \aap, 616, A17,
  \dodoi{10.1051/0004-6361/201833234}

\bibitem[{{Bate}(1998)}]{Bate1998}
{Bate}, M.~R. 1998, \apjl, 508, L95, \dodoi{10.1086/311719}

\bibitem[{{Bate}(2018)}]{Bate2018}
---. 2018, \mnras, 475, 5618, \dodoi{10.1093/mnras/sty169}

\bibitem[{{Bate} \& {Bonnell}(1997)}]{Bate1997}
{Bate}, M.~R., \& {Bonnell}, I.~A. 1997, \mnras, 285, 33,
  \dodoi{10.1093/mnras/285.1.33}

\bibitem[{{Bellm} {et~al.}(2019){Bellm}, {Kulkarni}, {Barlow}, {Feindt},
  {Graham}, {Goobar}, {Kupfer}, {Ngeow}, {Nugent}, {Ofek}, {Prince}, {Riddle},
  {Walters}, \& {Ye}}]{Bellm2019a}
{Bellm}, E.~C., {Kulkarni}, S.~R., {Barlow}, T., {et~al.} 2019, \pasp, 131,
  068003, \dodoi{10.1088/1538-3873/ab0c2a}

\bibitem[{{Bressan} {et~al.}(2012){Bressan}, {Marigo}, {Girardi}, {Salasnich},
  {Dal Cero}, {Rubele}, \& {Nanni}}]{Bressan2012}
{Bressan}, A., {Marigo}, P., {Girardi}, L., {et~al.} 2012, \mnras, 427, 127,
  \dodoi{10.1111/j.1365-2966.2012.21948.x}

\bibitem[{{Cui} {et~al.}(2012){Cui}, {Zhao}, {Chu}, {Li}, {Li}, {Zhang}, {Su},
  {Yao}, {Wang}, {Xing}, {Li}, {Zhu}, {Wang}, {Gu}, {Luo}, {Xu}, {Zhang},
  {Liu}, {Zhang}, {Yang}, {Cao}, {Chen}, {Chen}, {Chen}, {Chen}, {Chu}, {Feng},
  {Gong}, {Hou}, {Hu}, {Hu}, {Hu}, {Jia}, {Jiang}, {Jiang}, {Jiang}, {Jin},
  {Li}, {Li}, {Li}, {Liu}, {Liu}, {Lu}, {Mao}, {Men}, {Qi}, {Qi}, {Shi},
  {Tang}, {Tao}, {Wang}, {Wang}, {Wang}, {Wang}, {Wang}, {Wang}, {Wang},
  {Wang}, {Wang}, {Wang}, {Wang}, {Wang}, {Xu}, {Xu}, {Yang}, {Yu}, {Yuan},
  {Yuan}, {Zhai}, {Zhang}, {Zhang}, {Zhang}, {Zhao}, {Zhou}, {Zhou}, {Zhu}, \&
  {Zou}}]{Cui2012}
{Cui}, X.-Q., {Zhao}, Y.-H., {Chu}, Y.-Q., {et~al.} 2012, Research in Astronomy
  and Astrophysics, 12, 1197, \dodoi{10.1088/1674-4527/12/9/003}

\bibitem[{{Duch{\^e}ne} \& {Kraus}(2013)}]{Duchene2013}
{Duch{\^e}ne}, G., \& {Kraus}, A. 2013, \araa, 51, 269,
  \dodoi{10.1146/annurev-astro-081710-102602}

\bibitem[{{Duquennoy} \& {Mayor}(1991)}]{Duquennoy1991}
{Duquennoy}, A., \& {Mayor}, M. 1991, \aap, 500, 337

\bibitem[{{El-Badry} {et~al.}(2018){El-Badry}, {Ting}, {Rix}, {Quataert},
  {Weisz}, {Cargile}, {Conroy}, {Hogg}, {Bergemann}, \& {Liu}}]{El-Badry2018}
{El-Badry}, K., {Ting}, Y.-S., {Rix}, H.-W., {et~al.} 2018, \mnras, 476, 528,
  \dodoi{10.1093/mnras/sty240}

\bibitem[{{Flewelling}(2016)}]{Flewelling2016}
{Flewelling}, H. 2016, Commmunications of the Konkoly Observatory Hungary, 105,
  77

\bibitem[{{Gaia Collaboration} {et~al.}(2016){Gaia Collaboration}, {Prusti},
  {de Bruijne}, {Brown}, {Vallenari}, {Babusiaux}, {Bailer-Jones}, {Bastian},
  {Biermann}, {Evans}, {Eyer}, {Jansen}, {Jordi}, {Klioner}, {Lammers},
  {Lindegren}, {Luri}, {Mignard}, {Milligan}, {Panem}, {Poinsignon},
  {Pourbaix}, {Randich}, {Sarri}, {Sartoretti}, {Siddiqui}, {Soubiran},
  {Valette}, {van Leeuwen}, {Walton}, {Aerts}, {Arenou}, {Cropper}, {Drimmel},
  {H{\o}g}, {Katz}, {Lattanzi}, {O'Mullane}, {Grebel}, {Holland}, {Huc},
  {Passot}, {Bramante}, {Cacciari}, {Casta{\~n}eda}, {Chaoul}, {Cheek}, {De
  Angeli}, {Fabricius}, {Guerra}, {Hern{\'a}ndez}, {Jean-Antoine-Piccolo},
  {Masana}, {Messineo}, {Mowlavi}, {Nienartowicz}, {Ord{\'o}{\~n}ez-Blanco},
  {Panuzzo}, {Portell}, {Richards}, {Riello}, {Seabroke}, {Tanga},
  {Th{\'e}venin}, {Torra}, {Els}, {Gracia-Abril}, {Comoretto},
  {Garcia-Reinaldos}, {Lock}, {Mercier}, {Altmann}, {Andrae}, {Astraatmadja},
  {Bellas-Velidis}, {Benson}, {Berthier}, {Blomme}, {Busso}, {Carry},
  {Cellino}, {Clementini}, {Cowell}, {Creevey}, {Cuypers}, {Davidson}, {De
  Ridder}, {de Torres}, {Delchambre}, {Dell'Oro}, {Ducourant}, {Fr{\'e}mat},
  {Garc{\'\i}a-Torres}, {Gosset}, {Halbwachs}, {Hambly}, {Harrison}, {Hauser},
  {Hestroffer}, {Hodgkin}, {Huckle}, {Hutton}, {Jasniewicz}, {Jordan},
  {Kontizas}, {Korn}, {Lanzafame}, {Manteiga}, {Moitinho}, {Muinonen},
  {Osinde}, {Pancino}, {Pauwels}, {Petit}, {Recio-Blanco}, {Robin}, {Sarro},
  {Siopis}, {Smith}, {Smith}, {Sozzetti}, {Thuillot}, {van Reeven}, {Viala},
  {Abbas}, {Abreu Aramburu}, {Accart}, {Aguado}, {Allan}, {Allasia},
  {Altavilla}, {{\'A}lvarez}, {Alves}, {Anderson}, {Andrei}, {Anglada Varela},
  {Antiche}, {Antoja}, {Ant{\'o}n}, {Arcay}, {Atzei}, {Ayache}, {Bach},
  {Baker}, {Balaguer-N{\'u}{\~n}ez}, {Barache}, {Barata}, {Barbier}, {Barblan},
  {Baroni}, {Barrado y Navascu{\'e}s}, {Barros}, {Barstow}, {Becciani},
  {Bellazzini}, {Bellei}, {Bello Garc{\'\i}a}, {Belokurov}, {Bendjoya},
  {Berihuete}, {Bianchi}, {Bienaym{\'e}}, {Billebaud}, {Blagorodnova},
  {Blanco-Cuaresma}, {Boch}, {Bombrun}, {Borrachero}, {Bouquillon}, {Bourda},
  {Bouy}, {Bragaglia}, {Breddels}, {Brouillet}, {Br{\"u}semeister},
  {Bucciarelli}, {Budnik}, {Burgess}, {Burgon}, {Burlacu}, {Busonero}, {Buzzi},
  {Caffau}, {Cambras}, {Campbell}, {Cancelliere}, {Cantat-Gaudin}, {Carlucci},
  {Carrasco}, {Castellani}, {Charlot}, {Charnas}, {Charvet}, {Chassat},
  {Chiavassa}, {Clotet}, {Cocozza}, {Collins}, {Collins}, {Costigan}, {Crifo},
  {Cross}, {Crosta}, {Crowley}, {Dafonte}, {Damerdji}, {Dapergolas}, {David},
  {David}, {De Cat}, {de Felice}, {de Laverny}, {De Luise}, {De March}, {de
  Martino}, {de Souza}, {Debosscher}, {del Pozo}, {Delbo}, {Delgado},
  {Delgado}, {di Marco}, {Di Matteo}, {Diakite}, {Distefano}, {Dolding}, {Dos
  Anjos}, {Drazinos}, {Dur{\'a}n}, {Dzigan}, {Ecale}, {Edvardsson}, {Enke},
  {Erdmann}, {Escolar}, {Espina}, {Evans}, {Eynard Bontemps}, {Fabre},
  {Fabrizio}, {Faigler}, {Falc{\~a}o}, {Farr{\`a}s Casas}, {Faye}, {Federici},
  {Fedorets}, {Fern{\'a}ndez-Hern{\'a}ndez}, {Fernique}, {Fienga}, {Figueras},
  {Filippi}, {Findeisen}, {Fonti}, {Fouesneau}, {Fraile}, {Fraser}, {Fuchs},
  {Furnell}, {Gai}, {Galleti}, {Galluccio}, {Garabato}, {Garc{\'\i}a-Sedano},
  {Gar{\'e}}, {Garofalo}, {Garralda}, {Gavras}, {Gerssen}, {Geyer}, {Gilmore},
  {Girona}, {Giuffrida}, {Gomes}, {Gonz{\'a}lez-Marcos},
  {Gonz{\'a}lez-N{\'u}{\~n}ez}, {Gonz{\'a}lez-Vidal}, {Granvik}, {Guerrier},
  {Guillout}, {Guiraud}, {G{\'u}rpide}, {Guti{\'e}rrez-S{\'a}nchez}, {Guy},
  {Haigron}, {Hatzidimitriou}, {Haywood}, {Heiter}, {Helmi}, {Hobbs},
  {Hofmann}, {Holl}, {Holland }, {Hunt}, {Hypki}, {Icardi}, {Irwin}, {Jevardat
  de Fombelle}, {Jofr{\'e}}, {Jonker}, {Jorissen}, {Julbe}, {Karampelas},
  {Kochoska}, {Kohley}, {Kolenberg}, {Kontizas}, {Koposov}, {Kordopatis},
  {Koubsky}, {Kowalczyk}, {Krone-Martins}, {Kudryashova}, {Kull}, {Bachchan},
  {Lacoste-Seris}, {Lanza}, {Lavigne}, {Le Poncin-Lafitte}, {Lebreton},
  {Lebzelter}, {Leccia}, {Leclerc}, {Lecoeur-Taibi}, {Lemaitre}, {Lenhardt},
  {Leroux}, {Liao}, {Licata}, {Lindstr{\o}m}, {Lister}, {Livanou}, {Lobel},
  {L{\"o}ffler}, {L{\'o}pez}, {Lopez-Lozano}, {Lorenz}, {Loureiro},
  {MacDonald}, {Magalh{\~a}es Fernandes}, {Managau}, {Mann}, {Mantelet},
  {Marchal}, {Marchant}, {Marconi}, {Marie}, {Marinoni}, {Marrese},
  {Marschalk{\'o}}, {Marshall}, {Mart{\'\i}n-Fleitas}, {Martino}, {Mary},
  {Matijevi{\v{c}}}, {Mazeh}, {McMillan}, {Messina}, {Mestre}, {Michalik},
  {Millar}, {Miranda}, {Molina}, {Molinaro}, {Molinaro}, {Moln{\'a}r},
  {Moniez}, {Montegriffo}, {Monteiro}, {Mor}, {Mora}, {Morbidelli}, {Morel},
  {Morgenthaler}, {Morley}, {Morris}, {Mulone}, {Muraveva}, {Musella},
  {Narbonne}, {Nelemans}, {Nicastro}, {Noval}, {Ord{\'e}novic},
  {Ordieres-Mer{\'e}}, {Osborne}, {Pagani}, {Pagano}, {Pailler}, {Palacin},
  {Palaversa}, {Parsons}, {Paulsen}, {Pecoraro}, {Pedrosa}, {Pentik{\"a}inen},
  {Pereira}, {Pichon}, {Piersimoni}, {Pineau}, {Plachy}, {Plum}, {Poujoulet},
  {Pr{\v{s}}a}, {Pulone}, {Ragaini}, {Rago}, {Rambaux}, {Ramos-Lerate},
  {Ranalli}, {Rauw}, {Read}, {Regibo}, {Renk}, {Reyl{\'e}}, {Ribeiro},
  {Rimoldini}, {Ripepi}, {Riva}, {Rixon}, {Roelens}, {Romero-G{\'o}mez},
  {Rowell}, {Royer}, {Rudolph}, {Ruiz-Dern}, {Sadowski}, {Sagrist{\`a}
  Sell{\'e}s}, {Sahlmann}, {Salgado}, {Salguero}, {Sarasso}, {Savietto},
  {Schnorhk}, {Schultheis}, {Sciacca}, {Segol}, {Segovia}, {Segransan},
  {Serpell}, {Shih}, {Smareglia}, {Smart}, {Smith}, {Solano}, {Solitro},
  {Sordo}, {Soria Nieto}, {Souchay}, {Spagna}, {Spoto}, {Stampa}, {Steele},
  {Steidelm{\"u}ller}, {Stephenson}, {Stoev}, {Suess}, {S{\"u}veges}, {Surdej},
  {Szabados}, {Szegedi-Elek}, {Tapiador}, {Taris}, {Tauran}, {Taylor},
  {Teixeira}, {Terrett}, {Tingley}, {Trager}, {Turon}, {Ulla}, {Utrilla},
  {Valentini}, {van Elteren}, {Van Hemelryck}, {van Leeuwen}, {Varadi},
  {Vecchiato}, {Veljanoski}, {Via}, {Vicente}, {Vogt}, {Voss}, {Votruba},
  {Voutsinas}, {Walmsley}, {Weiler}, {Weingrill}, {Werner}, {Wevers},
  {Whitehead}, {Wyrzykowski}, {Yoldas}, {{\v{Z}}erjal}, {Zucker}, {Zurbach},
  {Zwitter}, {Alecu}, {Allen}, {Allende Prieto}, {Amorim},
  {Anglada-Escud{\'e}}, {Arsenijevic}, {Azaz}, {Balm}, {Beck}, {Bernstein},
  {Bigot}, {Bijaoui}, {Blasco}, {Bonfigli}, {Bono}, {Boudreault}, {Bressan},
  {Brown}, {Brunet}, {Bunclark}, {Buonanno}, {Butkevich}, {Carret}, {Carrion},
  {Chemin}, {Ch{\'e}reau}, {Corcione}, {Darmigny}, {de Boer}, {de Teodoro}, {de
  Zeeuw}, {Delle Luche}, {Domingues}, {Dubath}, {Fodor}, {Fr{\'e}zouls},
  {Fries}, {Fustes}, {Fyfe}, {Gallardo}, {Gallegos}, {Gardiol}, {Gebran},
  {Gomboc}, {G{\'o}mez}, {Grux}, {Gueguen}, {Heyrovsky}, {Hoar}, {Iannicola},
  {Isasi Parache}, {Janotto}, {Joliet}, {Jonckheere}, {Keil}, {Kim},
  {Klagyivik}, {Klar}, {Knude}, {Kochukhov}, {Kolka}, {Kos}, {Kutka}, {Lainey},
  {LeBouquin}, {Liu}, {Loreggia}, {Makarov}, {Marseille}, {Martayan},
  {Martinez-Rubi}, {Massart}, {Meynadier}, {Mignot}, {Munari}, {Nguyen},
  {Nordlander}, {Ocvirk}, {O'Flaherty}, {Olias Sanz}, {Ortiz}, {Osorio},
  {Oszkiewicz}, {Ouzounis}, {Palmer}, {Park}, {Pasquato}, {Peltzer}, {Peralta},
  {P{\'e}turaud}, {Pieniluoma}, {Pigozzi}, {Poels}, {Prat}, {Prod'homme},
  {Raison}, {Rebordao}, {Risquez}, {Rocca-Volmerange}, {Rosen}, {Ruiz-Fuertes},
  {Russo}, {Sembay}, {Serraller Vizcaino}, {Short}, {Siebert}, {Silva},
  {Sinachopoulos}, {Slezak}, {Soffel}, {Sosnowska}, {Strai{\v{z}}ys}, {ter
  Linden}, {Terrell}, {Theil}, {Tiede}, {Troisi}, {Tsalmantza}, {Tur},
  {Vaccari}, {Vachier}, {Valles}, {Van Hamme}, {Veltz}, {Virtanen}, {Wallut},
  {Wichmann}, {Wilkinson}, {Ziaeepour}, \& {Zschocke}}]{GaiaCollaboration2016}
{Gaia Collaboration}, {Prusti}, T., {de Bruijne}, J.~H.~J., {et~al.} 2016,
  \aap, 595, A1, \dodoi{10.1051/0004-6361/201629272}

\bibitem[{{Gaia Collaboration} {et~al.}(2018){Gaia Collaboration}, {Brown},
  {Vallenari}, {Prusti}, {de Bruijne}, {Babusiaux}, {Bailer-Jones}, {Biermann},
  {Evans}, {Eyer}, {Jansen}, {Jordi}, {Klioner}, {Lammers}, {Lindegren},
  {Luri}, {Mignard}, {Panem}, {Pourbaix}, {Randich}, {Sartoretti}, {Siddiqui},
  {Soubiran}, {van Leeuwen}, {Walton}, {Arenou}, {Bastian}, {Cropper},
  {Drimmel}, {Katz}, {Lattanzi}, {Bakker}, {Cacciari}, {Casta{\~n}eda},
  {Chaoul}, {Cheek}, {De Angeli}, {Fabricius}, {Guerra}, {Holl}, {Masana},
  {Messineo}, {Mowlavi}, {Nienartowicz}, {Panuzzo}, {Portell}, {Riello},
  {Seabroke}, {Tanga}, {Th{\'e}venin}, {Gracia-Abril}, {Comoretto},
  {Garcia-Reinaldos}, {Teyssier}, {Altmann}, {Andrae}, {Audard},
  {Bellas-Velidis}, {Benson}, {Berthier}, {Blomme}, {Burgess}, {Busso},
  {Carry}, {Cellino}, {Clementini}, {Clotet}, {Creevey}, {Davidson}, {De
  Ridder}, {Delchambre}, {Dell'Oro}, {Ducourant},
  {Fern{\'a}ndez-Hern{\'a}ndez}, {Fouesneau}, {Fr{\'e}mat}, {Galluccio},
  {Garc{\'\i}a-Torres}, {Gonz{\'a}lez-N{\'u}{\~n}ez}, {Gonz{\'a}lez-Vidal},
  {Gosset}, {Guy}, {Halbwachs}, {Hambly}, {Harrison}, {Hern{\'a}ndez},
  {Hestroffer}, {Hodgkin}, {Hutton}, {Jasniewicz}, {Jean-Antoine-Piccolo},
  {Jordan}, {Korn}, {Krone-Martins}, {Lanzafame}, {Lebzelter}, {L{\"o}ffler},
  {Manteiga}, {Marrese}, {Mart{\'\i}n-Fleitas}, {Moitinho}, {Mora}, {Muinonen},
  {Osinde}, {Pancino}, {Pauwels}, {Petit}, {Recio-Blanco}, {Richards},
  {Rimoldini}, {Robin}, {Sarro}, {Siopis}, {Smith}, {Sozzetti}, {S{\"u}veges},
  {Torra}, {van Reeven}, {Abbas}, {Abreu Aramburu}, {Accart}, {Aerts},
  {Altavilla}, {{\'A}lvarez}, {Alvarez}, {Alves}, {Anderson}, {Andrei},
  {Anglada Varela}, {Antiche}, {Antoja}, {Arcay}, {Astraatmadja}, {Bach},
  {Baker}, {Balaguer-N{\'u}{\~n}ez}, {Balm}, {Barache}, {Barata}, {Barbato},
  {Barblan}, {Barklem}, {Barrado}, {Barros}, {Barstow}, {Bartholom{\'e}
  Mu{\~n}oz}, {Bassilana}, {Becciani}, {Bellazzini}, {Berihuete}, {Bertone},
  {Bianchi}, {Bienaym{\'e}}, {Blanco-Cuaresma}, {Boch}, {Boeche}, {Bombrun},
  {Borrachero}, {Bossini}, {Bouquillon}, {Bourda}, {Bragaglia}, {Bramante},
  {Breddels}, {Bressan}, {Brouillet}, {Br{\"u}semeister}, {Brugaletta},
  {Bucciarelli}, {Burlacu}, {Busonero}, {Butkevich}, {Buzzi}, {Caffau},
  {Cancelliere}, {Cannizzaro}, {Cantat-Gaudin}, {Carballo}, {Carlucci},
  {Carrasco}, {Casamiquela}, {Castellani}, {Castro-Ginard}, {Charlot},
  {Chemin}, {Chiavassa}, {Cocozza}, {Costigan}, {Cowell}, {Crifo}, {Crosta},
  {Crowley}, {Cuypers}, {Dafonte}, {Damerdji}, {Dapergolas}, {David}, {David},
  {de Laverny}, {De Luise}, {De March}, {de Martino}, {de Souza}, {de Torres},
  {Debosscher}, {del Pozo}, {Delbo}, {Delgado}, {Delgado}, {Di Matteo},
  {Diakite}, {Diener}, {Distefano}, {Dolding}, {Drazinos}, {Dur{\'a}n},
  {Edvardsson}, {Enke}, {Eriksson}, {Esquej}, {Eynard Bontemps}, {Fabre},
  {Fabrizio}, {Faigler}, {Falc{\~a}o}, {Farr{\`a}s Casas}, {Federici},
  {Fedorets}, {Fernique}, {Figueras}, {Filippi}, {Findeisen}, {Fonti},
  {Fraile}, {Fraser}, {Fr{\'e}zouls}, {Gai}, {Galleti}, {Garabato},
  {Garc{\'\i}a-Sedano}, {Garofalo}, {Garralda}, {Gavel}, {Gavras}, {Gerssen},
  {Geyer}, {Giacobbe}, {Gilmore}, {Girona}, {Giuffrida}, {Glass}, {Gomes},
  {Granvik}, {Gueguen}, {Guerrier}, {Guiraud}, {Guti{\'e}rrez-S{\'a}nchez},
  {Haigron}, {Hatzidimitriou}, {Hauser}, {Haywood}, {Heiter}, {Helmi}, {Heu},
  {Hilger}, {Hobbs}, {Hofmann}, {Holland}, {Huckle}, {Hypki}, {Icardi},
  {Jan{\ss}en}, {Jevardat de Fombelle}, {Jonker}, {Juh{\'a}sz}, {Julbe},
  {Karampelas}, {Kewley}, {Klar}, {Kochoska}, {Kohley}, {Kolenberg},
  {Kontizas}, {Kontizas}, {Koposov}, {Kordopatis}, {Kostrzewa-Rutkowska},
  {Koubsky}, {Lambert}, {Lanza}, {Lasne}, {Lavigne}, {Le Fustec}, {Le
  Poncin-Lafitte}, {Lebreton}, {Leccia}, {Leclerc}, {Lecoeur-Taibi},
  {Lenhardt}, {Leroux}, {Liao}, {Licata}, {Lindstr{\o}m}, {Lister}, {Livanou},
  {Lobel}, {L{\'o}pez}, {Managau}, {Mann}, {Mantelet}, {Marchal}, {Marchant},
  {Marconi}, {Marinoni}, {Marschalk{\'o}}, {Marshall}, {Martino}, {Marton},
  {Mary}, {Massari}, {Matijevi{\v{c}}}, {Mazeh}, {McMillan}, {Messina},
  {Michalik}, {Millar}, {Molina}, {Molinaro}, {Moln{\'a}r}, {Montegriffo},
  {Mor}, {Morbidelli}, {Morel}, {Morris}, {Mulone}, {Muraveva}, {Musella},
  {Nelemans}, {Nicastro}, {Noval}, {O'Mullane}, {Ord{\'e}novic},
  {Ord{\'o}{\~n}ez-Blanco}, {Osborne}, {Pagani}, {Pagano}, {Pailler},
  {Palacin}, {Palaversa}, {Panahi}, {Pawlak}, {Piersimoni}, {Pineau}, {Plachy},
  {Plum}, {Poggio}, {Poujoulet}, {Pr{\v{s}}a}, {Pulone}, {Racero}, {Ragaini},
  {Rambaux}, {Ramos-Lerate}, {Regibo}, {Reyl{\'e}}, {Riclet}, {Ripepi}, {Riva},
  {Rivard}, {Rixon}, {Roegiers}, {Roelens}, {Romero-G{\'o}mez}, {Rowell},
  {Royer}, {Ruiz-Dern}, {Sadowski}, {Sagrist{\`a} Sell{\'e}s}, {Sahlmann},
  {Salgado}, {Salguero}, {Sanna}, {Santana-Ros}, {Sarasso}, {Savietto},
  {Schultheis}, {Sciacca}, {Segol}, {Segovia}, {S{\'e}gransan}, {Shih},
  {Siltala}, {Silva}, {Smart}, {Smith}, {Solano}, {Solitro}, {Sordo}, {Soria
  Nieto}, {Souchay}, {Spagna}, {Spoto}, {Stampa}, {Steele},
  {Steidelm{\"u}ller}, {Stephenson}, {Stoev}, {Suess}, {Surdej}, {Szabados},
  {Szegedi-Elek}, {Tapiador}, {Taris}, {Tauran}, {Taylor}, {Teixeira},
  {Terrett}, {Teyssand ier}, {Thuillot}, {Titarenko}, {Torra Clotet}, {Turon},
  {Ulla}, {Utrilla}, {Uzzi}, {Vaillant}, {Valentini}, {Valette}, {van Elteren},
  {Van Hemelryck}, {van Leeuwen}, {Vaschetto}, {Vecchiato}, {Veljanoski},
  {Viala}, {Vicente}, {Vogt}, {von Essen}, {Voss}, {Votruba}, {Voutsinas},
  {Walmsley}, {Weiler}, {Wertz}, {Wevers}, {Wyrzykowski}, {Yoldas},
  {{\v{Z}}erjal}, {Ziaeepour}, {Zorec}, {Zschocke}, {Zucker}, {Zurbach}, \&
  {Zwitter}}]{GaiaCollaboration2018}
{Gaia Collaboration}, {Brown}, A.~G.~A., {Vallenari}, A., {et~al.} 2018, \aap,
  616, A1, \dodoi{10.1051/0004-6361/201833051}

\bibitem[{{Ge} {et~al.}(2010){Ge}, {Hjellming}, {Webbink}, {Chen}, \&
  {Han}}]{Ge2010}
{Ge}, H., {Hjellming}, M.~S., {Webbink}, R.~F., {Chen}, X., \& {Han}, Z. 2010,
  \apj, 717, 724, \dodoi{10.1088/0004-637X/717/2/724}

\bibitem[{{Ge} {et~al.}(2015){Ge}, {Webbink}, {Chen}, \& {Han}}]{Ge2015}
{Ge}, H., {Webbink}, R.~F., {Chen}, X., \& {Han}, Z. 2015, \apj, 812, 40,
  \dodoi{10.1088/0004-637X/812/1/40}

\bibitem[{{Ge} {et~al.}(2020{\natexlab{a}}){Ge}, {Webbink}, {Chen}, \&
  {Han}}]{Ge2020b}
---. 2020{\natexlab{a}}, \apj, 899, 132, \dodoi{10.3847/1538-4357/aba7b7}

\bibitem[{{Ge} {et~al.}(2020{\natexlab{b}}){Ge}, {Webbink}, {Chen}, \&
  {Han}}]{Ge2020}
---. 2020{\natexlab{b}}, \apj, 899, 132, \dodoi{10.3847/1538-4357/aba7b7}

\bibitem[{{Ge} {et~al.}(2020{\natexlab{c}}){Ge}, {Webbink}, \& {Han}}]{Ge2020a}
{Ge}, H., {Webbink}, R.~F., \& {Han}, Z. 2020{\natexlab{c}}, \apjs, 249, 9,
  \dodoi{10.3847/1538-4365/ab98f6}

\bibitem[{{Graham} {et~al.}(2019){Graham}, {Kulkarni}, {Bellm}, {Adams},
  {Barbarino}, {Blagorodnova}, {Bodewits}, {Bolin}, {Brady}, {Cenko}, {Chang},
  {Coughlin}, {De}, {Eadie}, {Farnham}, {Feindt}, {Franckowiak}, {Fremling},
  {Gezari}, {Ghosh}, {Goldstein}, {Golkhou}, {Goobar}, {Ho}, {Huppenkothen},
  {Ivezi{\'c}}, {Jones}, {Juric}, {Kaplan}, {Kasliwal}, {Kelley}, {Kupfer},
  {Lee}, {Lin}, {Lunnan}, {Mahabal}, {Miller}, {Ngeow}, {Nugent}, {Ofek},
  {Prince}, {Rauch}, {van Roestel}, {Schulze}, {Singer}, {Sollerman}, {Taddia},
  {Yan}, {Ye}, {Yu}, {Barlow}, {Bauer}, {Beck}, {Belicki}, {Biswas}, {Brinnel},
  {Brooke}, {Bue}, {Bulla}, {Burruss}, {Connolly}, {Cromer}, {Cunningham},
  {Dekany}, {Delacroix}, {Desai}, {Duev}, {Feeney}, {Flynn}, {Frederick},
  {Gal-Yam}, {Giomi}, {Groom}, {Hacopians}, {Hale}, {Helou}, {Henning},
  {Hover}, {Hillenbrand}, {Howell}, {Hung}, {Imel}, {Ip}, {Jackson}, {Kaspi},
  {Kaye}, {Kowalski}, {Kramer}, {Kuhn}, {Landry}, {Laher}, {Mao}, {Masci},
  {Monkewitz}, {Murphy}, {Nordin}, {Patterson}, {Penprase}, {Porter},
  {Rebbapragada}, {Reiley}, {Riddle}, {Rigault}, {Rodriguez}, {Rusholme}, {van
  Santen}, {Shupe}, {Smith}, {Soumagnac}, {Stein}, {Surace}, {Szkody}, {Terek},
  {Van Sistine}, {van Velzen}, {Vestrand}, {Walters}, {Ward}, {Zhang}, \&
  {Zolkower}}]{Graham2019}
{Graham}, M.~J., {Kulkarni}, S.~R., {Bellm}, E.~C., {et~al.} 2019, \pasp, 131,
  078001, \dodoi{10.1088/1538-3873/ab006c}

\bibitem[{{Gray} \& {Corbally}(1994)}]{Gray1994}
{Gray}, R.~O., \& {Corbally}, C.~J. 1994, \aj, 107, 742, \dodoi{10.1086/116893}

\bibitem[{{Han}(1998)}]{Han1998}
{Han}, Z. 1998, \mnras, 296, 1019, \dodoi{10.1046/j.1365-8711.1998.01475.x}

\bibitem[{{Han} {et~al.}(2002){Han}, {Podsiadlowski}, {Maxted}, {Marsh}, \&
  {Ivanova}}]{Han2002}
{Han}, Z., {Podsiadlowski}, P., {Maxted}, P.~F.~L., {Marsh}, T.~R., \&
  {Ivanova}, N. 2002, \mnras, 336, 449,
  \dodoi{10.1046/j.1365-8711.2002.05752.x}

\bibitem[{{Han} {et~al.}(2020){Han}, {Ge}, {Chen}, \& {Chen}}]{Han2020}
{Han}, Z.-W., {Ge}, H.-W., {Chen}, X.-F., \& {Chen}, H.-L. 2020, Research in
  Astronomy and Astrophysics, 20, 161, \dodoi{10.1088/1674-4527/20/10/161}

\bibitem[{{Hjellming} \& {Webbink}(1987)}]{Hjellming1987}
{Hjellming}, M.~S., \& {Webbink}, R.~F. 1987, \apj, 318, 794,
  \dodoi{10.1086/165412}

\bibitem[{{Holmberg} {et~al.}(2009){Holmberg}, {Nordstr{\"o}m}, \&
  {Andersen}}]{Holmberg2009}
{Holmberg}, J., {Nordstr{\"o}m}, B., \& {Andersen}, J. 2009, \aap, 501, 941,
  \dodoi{10.1051/0004-6361/200811191}

\bibitem[{{Jones} \& {Boffin}(2017)}]{Jones2017}
{Jones}, D., \& {Boffin}, H. M.~J. 2017, Nature Astronomy, 1, 0117,
  \dodoi{10.1038/s41550-017-0117}

\bibitem[{{Koleva} {et~al.}(2009){Koleva}, {Prugniel}, {Bouchard}, \&
  {Wu}}]{Koleva2009}
{Koleva}, M., {Prugniel}, P., {Bouchard}, A., \& {Wu}, Y. 2009, \aap, 501,
  1269, \dodoi{10.1051/0004-6361/200811467}

\bibitem[{{Kounkel}(2021)}]{Kounkel2021}
{Kounkel}, M. 2021, {APOGEE SB2s}, V1.1,  Zenodo,
  \dodoi{10.5281/zenodo.5068312}

\bibitem[{{Kouwenhoven} {et~al.}(2009){Kouwenhoven}, {Brown}, {Goodwin},
  {Portegies Zwart}, \& {Kaper}}]{Kouwenhoven2009}
{Kouwenhoven}, M.~B.~N., {Brown}, A.~G.~A., {Goodwin}, S.~P., {Portegies
  Zwart}, S.~F., \& {Kaper}, L. 2009, \aap, 493, 979,
  \dodoi{10.1051/0004-6361:200810234}

\bibitem[{{Kouwenhoven} {et~al.}(2007){Kouwenhoven}, {Brown}, {Portegies
  Zwart}, \& {Kaper}}]{Kouwenhoven2007}
{Kouwenhoven}, M.~B.~N., {Brown}, A.~G.~A., {Portegies Zwart}, S.~F., \&
  {Kaper}, L. 2007, \aap, 474, 77, \dodoi{10.1051/0004-6361:20077719}

\bibitem[{{Kroupa}(1995{\natexlab{a}})}]{Kroupa1995b}
{Kroupa}, P. 1995{\natexlab{a}}, \mnras, 277, 1491,
  \dodoi{10.1093/mnras/277.4.1491}

\bibitem[{{Kroupa}(1995{\natexlab{b}})}]{Kroupa1995a}
---. 1995{\natexlab{b}}, \mnras, 277, 1507, \dodoi{10.1093/mnras/277.4.1507}

\bibitem[{{Kruckow} {et~al.}(2021){Kruckow}, {Neunteufel}, {Di Stefano}, {Gao},
  \& {Kobayashi}}]{Kruckow2021}
{Kruckow}, M.~U., {Neunteufel}, P.~G., {Di Stefano}, R., {Gao}, Y., \&
  {Kobayashi}, C. 2021, \apj, 920, 86, \dodoi{10.3847/1538-4357/ac13ac}

\bibitem[{{Kurucz}(1993)}]{Kurucz1993}
{Kurucz}, R.~L. 1993, {SYNTHE spectrum synthesis programs and line data}

\bibitem[{{Kurucz} {et~al.}(1984){Kurucz}, {Furenlid}, {Brault}, \&
  {Testerman}}]{Kurucz1984}
{Kurucz}, R.~L., {Furenlid}, I., {Brault}, J., \& {Testerman}, L. 1984, {Solar
  flux atlas from 296 to 1300 nm}

\bibitem[{{Li} {et~al.}(2021){Li}, {Shi}, {Yan}, {Fu}, {Li}, \&
  {Hou}}]{Lichunqian2021}
{Li}, C.-q., {Shi}, J.-r., {Yan}, H.-l., {et~al.} 2021, \apjs, 256, 31,
  \dodoi{10.3847/1538-4365/ac22a8}

\bibitem[{{Liu}(2019)}]{Liu2019}
{Liu}, C. 2019, \mnras, 490, 550, \dodoi{10.1093/mnras/stz2274}

\bibitem[{{Liu} {et~al.}(2020){Liu}, {Fu}, {Shi}, {Wu}, {Han}, {Chen}, {Dong},
  {Zhao}, {Chen}, {Zhang}, {Bai}, {Chen}, {Cui}, {Du}, {Hsia}, {Jiang}, {Hou},
  {Hou}, {Li}, {Li}, {Li}, {Liu}, {Liu}, {Luo}, {Ren}, {Tian}, {Tian}, {Wang},
  {Wu}, {Xie}, {Yan}, {Yang}, {Yu}, {Zhang}, {Zhang}, {Zhang}, {Zhang}, {Zhao},
  {Zhong}, {Zong}, \& {Zuo}}]{Liu2020}
{Liu}, C., {Fu}, J., {Shi}, J., {et~al.} 2020, arXiv e-prints,
  arXiv:2005.07210.
\newblock \doarXiv{2005.07210}

\bibitem[{{Luo} {et~al.}(2015){Luo}, {Zhao}, {Zhao}, {Deng}, {Liu}, {Jing},
  {Wang}, {Zhang}, {Shi}, {Cui}, {Chu}, {Li}, {Bai}, {Wu}, {Cai}, {Cao}, {Cao},
  {Carlin}, {Chen}, {Chen}, {Chen}, {Chen}, {Chen}, {Chen}, {Chen},
  {Christlieb}, {Chu}, {Cui}, {Dong}, {Du}, {Fan}, {Feng}, {Fu}, {Gao}, {Gong},
  {Gu}, {Guo}, {Han}, {He}, {Hou}, {Hou}, {Hou}, {Hu}, {Hu}, {Hu}, {Huo},
  {Jia}, {Jiang}, {Jiang}, {Jiang}, {Jin}, {Kong}, {Kong}, {Lei}, {Li}, {Li},
  {Li}, {Li}, {Li}, {Li}, {Li}, {Li}, {Li}, {Li}, {Li}, {Li}, {Liang}, {Lin},
  {Liu}, {Liu}, {Liu}, {Liu}, {Lu}, {Luo}, {Mao}, {Newberg}, {Ni}, {Qi}, {Qi},
  {Shen}, {Shi}, {Song}, {Song}, {Su}, {Su}, {Tang}, {Tao}, {Tian}, {Wang},
  {Wang}, {Wang}, {Wang}, {Wang}, {Wang}, {Wang}, {Wang}, {Wang}, {Wang},
  {Wang}, {Wang}, {Wang}, {Wang}, {Wang}, {Wang}, {Wang}, {Wang}, {Wang},
  {Wang}, {Wei}, {Wei}, {Wu}, {Wu}, {Wu}, {Wu}, {Xing}, {Xu}, {Xu}, {Xu},
  {Yan}, {Yang}, {Yang}, {Yang}, {Yang}, {Yao}, {Yu}, {Yuan}, {Yuan}, {Yuan},
  {Yuan}, {Zhai}, {Zhang}, {Zhang}, {Zhang}, {Zhang}, {Zhang}, {Zhang},
  {Zhang}, {Zhang}, {Zhao}, {Zhou}, {Zhou}, {Zhu}, {Zhu}, {Zou}, \&
  {Zuo}}]{Luo2015}
{Luo}, A.~L., {Zhao}, Y.-H., {Zhao}, G., {et~al.} 2015, Research in Astronomy
  and Astrophysics, 15, 1095, \dodoi{10.1088/1674-4527/15/8/002}

\bibitem[{{Majewski} {et~al.}(2017){Majewski}, {Schiavon}, {Frinchaboy},
  {Allende Prieto}, {Barkhouser}, {Bizyaev}, {Blank}, {Brunner}, {Burton},
  {Carrera}, {Chojnowski}, {Cunha}, {Epstein}, {Fitzgerald}, {Garc{\'\i}a
  P{\'e}rez}, {Hearty}, {Henderson}, {Holtzman}, {Johnson}, {Lam}, {Lawler},
  {Maseman}, {M{\'e}sz{\'a}ros}, {Nelson}, {Nguyen}, {Nidever}, {Pinsonneault},
  {Shetrone}, {Smee}, {Smith}, {Stolberg}, {Skrutskie}, {Walker}, {Wilson},
  {Zasowski}, {Anders}, {Basu}, {Beland}, {Blanton}, {Bovy}, {Brownstein},
  {Carlberg}, {Chaplin}, {Chiappini}, {Eisenstein}, {Elsworth}, {Feuillet},
  {Fleming}, {Galbraith-Frew}, {Garc{\'\i}a}, {Garc{\'\i}a-Hern{\'a}ndez},
  {Gillespie}, {Girardi}, {Gunn}, {Hasselquist}, {Hayden}, {Hekker}, {Ivans},
  {Kinemuchi}, {Klaene}, {Mahadevan}, {Mathur}, {Mosser}, {Muna}, {Munn},
  {Nichol}, {O'Connell}, {Parejko}, {Robin}, {Rocha-Pinto}, {Schultheis},
  {Serenelli}, {Shane}, {Silva Aguirre}, {Sobeck}, {Thompson}, {Troup},
  {Weinberg}, \& {Zamora}}]{Majewski2017}
{Majewski}, S.~R., {Schiavon}, R.~P., {Frinchaboy}, P.~M., {et~al.} 2017, \aj,
  154, 94, \dodoi{10.3847/1538-3881/aa784d}

\bibitem[{{Matijevi{\v{c}}} {et~al.}(2010){Matijevi{\v{c}}}, {Zwitter},
  {Munari}, {Bienaym{\'e}}, {Binney}, {Bland-Hawthorn}, {Boeche}, {Campbell},
  {Freeman}, {Gibson}, {Gilmore}, {Grebel}, {Helmi}, {Navarro}, {Parker},
  {Seabroke}, {Siebert}, {Siviero}, {Steinmetz}, {Watson}, {Williams}, \&
  {Wyse}}]{Matijevic2010}
{Matijevi{\v{c}}}, G., {Zwitter}, T., {Munari}, U., {et~al.} 2010, \aj, 140,
  184, \dodoi{10.1088/0004-6256/140/1/184}

\bibitem[{{McDonald} \& {Clarke}(1995)}]{McDonald1995}
{McDonald}, J.~M., \& {Clarke}, C.~J. 1995, \mnras, 275, 671,
  \dodoi{10.1093/mnras/275.3.671}

\bibitem[{{Merle} {et~al.}(2017){Merle}, {Van Eck}, {Jorissen}, {Van der
  Swaelmen}, {Masseron}, {Zwitter}, {Hatzidimitriou}, {Klutsch}, {Pourbaix},
  {Blomme}, {Worley}, {Sacco}, {Lewis}, {Abia}, {Traven}, {Sordo}, {Bragaglia},
  {Smiljanic}, {Pancino}, {Damiani}, {Hourihane}, {Gilmore}, {Randich},
  {Koposov}, {Casey}, {Morbidelli}, {Franciosini}, {Magrini}, {Jofre},
  {Costado}, {Jeffries}, {Bergemann}, {Lanzafame}, {Bayo}, {Carraro},
  {Flaccomio}, {Monaco}, \& {Zaggia}}]{Merle2017}
{Merle}, T., {Van Eck}, S., {Jorissen}, A., {et~al.} 2017, \aap, 608, A95,
  \dodoi{10.1051/0004-6361/201730442}

\bibitem[{{Moe} \& {Di Stefano}(2017)}]{Moe2017}
{Moe}, M., \& {Di Stefano}, R. 2017, \apjs, 230, 15,
  \dodoi{10.3847/1538-4365/aa6fb6}

\bibitem[{{Nordstr{\"o}m} {et~al.}(2004){Nordstr{\"o}m}, {Mayor}, {Andersen},
  {Holmberg}, {Pont}, {J{\o}rgensen}, {Olsen}, {Udry}, \&
  {Mowlavi}}]{Nordstrom2004}
{Nordstr{\"o}m}, B., {Mayor}, M., {Andersen}, J., {et~al.} 2004, \aap, 418,
  989, \dodoi{10.1051/0004-6361:20035959}

\bibitem[{{Offner} {et~al.}(2016){Offner}, {Dunham}, {Lee}, {Arce}, \&
  {Fielding}}]{Offner2016}
{Offner}, S. S.~R., {Dunham}, M.~M., {Lee}, K.~I., {Arce}, H.~G., \&
  {Fielding}, D.~B. 2016, \apjl, 827, L11, \dodoi{10.3847/2041-8205/827/1/L11}

\bibitem[{{Offner} {et~al.}(2022){Offner}, {Moe}, {Kratter}, {Sadavoy},
  {Jensen}, \& {Tobin}}]{Offner2022}
{Offner}, S. S.~R., {Moe}, M., {Kratter}, K.~M., {et~al.} 2022, arXiv e-prints,
  arXiv:2203.10066.
\newblock \doarXiv{2203.10066}

\bibitem[{{Pourbaix} {et~al.}(2004){Pourbaix}, {Tokovinin}, {Batten}, {Fekel},
  {Hartkopf}, {Levato}, {Morrell}, {Torres}, \& {Udry}}]{Pourbaix2004}
{Pourbaix}, D., {Tokovinin}, A.~A., {Batten}, A.~H., {et~al.} 2004, \aap, 424,
  727, \dodoi{10.1051/0004-6361:20041213}

\bibitem[{{Prugniel} \& {Soubiran}(2001)}]{Prugniel2001}
{Prugniel}, P., \& {Soubiran}, C. 2001, \aap, 369, 1048,
  \dodoi{10.1051/0004-6361:20010163}

\bibitem[{{Prugniel} {et~al.}(2007){Prugniel}, {Soubiran}, {Koleva}, \& {Le
  Borgne}}]{Prugniel2007}
{Prugniel}, P., {Soubiran}, C., {Koleva}, M., \& {Le Borgne}, D. 2007, arXiv
  e-prints, astro.
\newblock \doarXiv{astro-ph/0703658}

\bibitem[{{Raghavan} {et~al.}(2010){Raghavan}, {McAlister}, {Henry}, {Latham},
  {Marcy}, {Mason}, {Gies}, {White}, \& {ten Brummelaar}}]{Raghavan2010}
{Raghavan}, D., {McAlister}, H.~A., {Henry}, T.~J., {et~al.} 2010, \apjs, 190,
  1, \dodoi{10.1088/0067-0049/190/1/1}

\bibitem[{{Sana} {et~al.}(2012){Sana}, {de Mink}, {de Koter}, {Langer},
  {Evans}, {Gieles}, {Gosset}, {Izzard}, {Le Bouquin}, \&
  {Schneider}}]{Sana2012}
{Sana}, H., {de Mink}, S.~E., {de Koter}, A., {et~al.} 2012, Science, 337, 444,
  \dodoi{10.1126/science.1223344}

\bibitem[{{Shatsky} \& {Tokovinin}(2002)}]{Shatsky2002}
{Shatsky}, N., \& {Tokovinin}, A. 2002, \aap, 382, 92,
  \dodoi{10.1051/0004-6361:20011542}

\bibitem[{{Simon} \& {Obbie}(2009)}]{Simon2009}
{Simon}, M., \& {Obbie}, R.~C. 2009, \aj, 137, 3442,
  \dodoi{10.1088/0004-6256/137/2/3442}

\bibitem[{{Tokovinin} \& {Moe}(2020)}]{Tokovinin2020}
{Tokovinin}, A., \& {Moe}, M. 2020, \mnras, 491, 5158,
  \dodoi{10.1093/mnras/stz3299}

\bibitem[{{Tokovinin}(1992)}]{Tokovinin1992}
{Tokovinin}, A.~A. 1992, \aap, 256, 121

\bibitem[{{Tokovinin}(2000)}]{Tokovinin2000}
---. 2000, \aap, 360, 997

\bibitem[{{Tout}(1991)}]{Tout1991}
{Tout}, C.~A. 1991, \mnras, 250, 701, \dodoi{10.1093/mnras/250.4.701}

\bibitem[{{Traven} {et~al.}(2020){Traven}, {Feltzing}, {Merle}, {Van der
  Swaelmen}, {{\v{C}}otar}, {Church}, {Zwitter}, {Ting}, {Sahlholdt},
  {Asplund}, {Bland-Hawthorn}, {De Silva}, {Freeman}, {Martell}, {Sharma},
  {Zucker}, {Buder}, {Casey}, {D'Orazi}, {Kos}, {Lewis}, {Lin}, {Lind},
  {Simpson}, {Stello}, {Munari}, \& {Wittenmyer}}]{Traven2020}
{Traven}, G., {Feltzing}, S., {Merle}, T., {et~al.} 2020, \aap, 638, A145,
  \dodoi{10.1051/0004-6361/202037484}

\bibitem[{{Wang} {et~al.}(2020){Wang}, {Luo}, {Chen}, {Hou}, {Zhang}, {Zhao},
  {Li}, {Hou}, \& {LAMOST MRS Collaboration}}]{Wang2020}
{Wang}, R., {Luo}, A.~L., {Chen}, J.-J., {et~al.} 2020, \apj, 891, 23,
  \dodoi{10.3847/1538-4357/ab6dea}

\bibitem[{{Wu} {et~al.}(2014){Wu}, {Du}, {Luo}, {Zhao}, \& {Yuan}}]{Wu2014}
{Wu}, Y., {Du}, B., {Luo}, A., {Zhao}, Y., \& {Yuan}, H. 2014, in Statistical
  Challenges in 21st Century Cosmology, ed. A.~{Heavens}, J.-L. {Starck}, \&
  A.~{Krone-Martins}, Vol. 306, 340--342, \dodoi{10.1017/S1743921314010825}

\bibitem[{{Wu} {et~al.}(2011){Wu}, {Luo}, {Li}, {Shi}, {Prugniel}, {Liang},
  {Zhao}, {Zhang}, {Bai}, {Wei}, {Dong}, {Zhang}, \& {Chen}}]{Wu2011}
{Wu}, Y., {Luo}, A.~L., {Li}, H.-N., {et~al.} 2011, Research in Astronomy and
  Astrophysics, 11, 924, \dodoi{10.1088/1674-4527/11/8/006}

\bibitem[{{Zhang} {et~al.}(2020){Zhang}, {Liu}, \& {Deng}}]{Zhang2020}
{Zhang}, B., {Liu}, C., \& {Deng}, L.-C. 2020, \apjs, 246, 9,
  \dodoi{10.3847/1538-4365/ab55ef}

\bibitem[{{Zhang} {et~al.}(2021){Zhang}, {Li}, {Yang}, {Xiong}, {Fu}, {Liu},
  {Tian}, {Li}, {Wang}, {Liang}, {Zhou}, {Zong}, {Yang}, {Liu}, \&
  {Hou}}]{Zhang2021}
{Zhang}, B., {Li}, J., {Yang}, F., {et~al.} 2021, arXiv e-prints,
  arXiv:2105.11624.
\newblock \doarXiv{2105.11624}

\bibitem[{{Zhao} {et~al.}(2012){Zhao}, {Zhao}, {Chu}, {Jing}, \&
  {Deng}}]{Zhao2012}
{Zhao}, G., {Zhao}, Y.-H., {Chu}, Y.-Q., {Jing}, Y.-P., \& {Deng}, L.-C. 2012,
  Research in Astronomy and Astrophysics, 12, 723,
  \dodoi{10.1088/1674-4527/12/7/002}

\end{thebibliography}
\bibliographystyle{aasjournal}

\end{CJK}
\end{document}